\documentclass[fleqn,usenatbib]{mnras}

\usepackage{newtxtext,newtxmath}
\usepackage[T1]{fontenc}
\usepackage{ae,aecompl}

\usepackage{graphicx}	% Including figure files
\usepackage{amsmath}	% Advanced maths commands
\usepackage{mathrsfs}
\usepackage{blindtext}
\usepackage{rotating}
\usepackage{pdflscape}
\usepackage{adjustbox}
\usepackage{marvosym}
\usepackage{booktabs}
\usepackage{blindtext}
\newcommand{\LCDM}{$\Lambda$CDM\ }

\newcommand{\cMpc}{\: \mathrm{cMpc}}
\newcommand{\hMpc}{\: \mathrm{cMpc}/h}
\newcommand{\Msun}{\: \mathrm{M}_{\odot}}
\newcommand{\K}{\: \mathrm{K}}
\newcommand{\simba}{\textsc{Simba}}

\title[Effects of baryons on halos \& LSS in \textsc{Simba}]{How baryons affect halos and large-scale structure: a unified picture from the \textsc{Simba} simulation}
\author[D. Sorini et al.]{
Daniele Sorini,$^{1, 2}$\thanks{E-mail: sorini@roe.ac.uk}
Romeel Dav\'e,$^{1, 3, 4}$
Weiguang Cui$^{1, 5}$ \&
Sarah Appleby$^{1}$
\\
$^{1}$Institute for Astronomy, University of Edinburgh, Blackford Hill, Edinburgh, EH9 3HJ, United Kingdom\\
$^2$D\'epartement de Physique Th\'eorique, Universit\'e de Gen\`eve, 24 quai Ernest Ansermet, 1211 Gen\`eve 4, Switzerland\\
$^{3}$University of the Western Cape, Bellville, Cape Town 7535, South Africa\\
$^{4}$South African Astronomical Observatories, Observatory, Cape Town 7925, South Africa\\
$^{5}$Departamento de F\'isica Te\'{o}rica, M\'{o}dulo 15, Facultad de Ciencias, Universidad Aut\'{o}noma de Madrid, 28049 Madrid, Spain\\
}

\date{Accepted XXX. Received YYY; in original form ZZZ}

\pubyear{2021}

\begin{document}
\label{firstpage}
\pagerange{\pageref{firstpage}--\pageref{lastpage}}
\maketitle

\begin{abstract}
\noindent
Using the state-of-the-art suite of hydrodynamic simulations \simba, as well as its dark-matter-only counterpart, we study the impact of the presence of baryons and of different stellar/AGN feedback mechanisms on large-scale structure, halo density profiles, and on the abundance of different baryonic phases within halos and in the intergalactic medium (IGM). The unified picture that emerges from our analysis is that the main physical drivers shaping the distribution of matter at all scales are star formation-driven galactic outflows at $z>2$ for lower mass halos and AGN jets at $z<2$ in higher mass halos. Feedback suppresses the baryon mass function with time relative to the halo mass function, and it even impacts the halo mass function itself at the $\sim 20\%$ level, particularly evacuating the centres and enhancing dark matter just outside halos. At early epochs baryons pile up in the centres of halos, but by late epochs and particularly in massive systems gas has mostly been evacuated from within the inner halo. AGN jets are so efficient at such evacuation that at low redshifts the baryon fraction within $\sim 10^{12}-10^{13} \Msun$ halos is only $25\%$ of the cosmic baryon fraction, mostly in stars. The baryon fraction enclosed in a sphere around such halos approaches the cosmic value $\Omega_{\rm b}/\Omega_{\rm m}$ only at $10-20$ virial radii. As a result, 87\% of the baryonic mass in the Universe lies in the IGM at $z=0$, with 67\% being in the form of warm-hot IGM ($T>10^5 \K$). 
\end{abstract}


%%%%%%%%%%%%%%%%%%%%%%%%%%%%%%%%%%%%%%%%%%%%%%%%%%

%%%%%%%%%%%%%%%%% BODY OF PAPER %%%%%%%%%%%%%%%%%%

\section{Introduction}

Understanding the emergence of galaxies from the growth of structures in the Universe is one of the primary goals of cosmological research. Within the standard $\Lambda$CDM paradigm, it is well established that dark matter halos form via hierarchical merging. Such process can be well described analytically \citep{LaceyCole1993}, and is validated by the results of large N-body cosmological simulations \citep[e.g.][]{Springel_2005, Klypin_2011, Angulo_2012, Fosalba_2015}. On the other hand, unveiling the details of the astrophysical processes that govern the build up and evolution of galaxies within dark matter halos proves to be much more challenging. 

While analytic models can provide valuable insight in this respect and succeed at broadly reproducing observations of the overall star formation history \citep[e.g.][]{White_1991, HS03, Rasera_2006, Dave_2012, Behroozi_2013_model, Moster_2018, Behroozi_2019, Ikea, Salcido_2018, Salcido_2020, Fukugita_2021, SP21}, they often do so by sacrificing physical realism to some degree. Because of the complex and interconnected nature of the underlying physical processes, hydrodynamic simulations represent one of the most favoured tools to model galaxy formation in a cosmological context. Though, this does not come without its difficulties either. While cosmological simulations aim at simultaneously reproducing the large-scale structure of the Universe and the inner structure of galaxies, they are of course limited by their finite resolution and computational cost. For this reason, it becomes necessary to characterise sub-grid processes such as the feedback of stellar winds and active galactic nuclei (AGN) on star formation via numerical prescriptions that can vary from code to code \citep[see][for a review]{feedback_review}. Several several hydrodynamic simulations \citep[e.g.][]{OWLS, Almgren_2013, Enzo_Bryan2014, Dubois_2014, FIRE2014, Illustris_V2014, Lukic_2015, EAGLE_Schaye2015, Mufasa, McCarty_2017, IllustrisTNG2018, Simba} manage to produce realistic galaxy populations despite their different feedback implementations. It is therefore of great interest to identify observables capable of discriminating among the predictions of different feedback models, and to closely examine the effects of such models on a variety of aspects of structure and galaxy formation.

Even though feedback mechanisms were originally introduced as an explanation for the observed quenching of star formation at $z<2$ \citep[see reviews by][]{Madau_rev, feedback_review}, they have other notable consequences too. Feedback processes - and indeed the mere presence of baryons - can affect the overall distribution of matter in the Universe and the internal structure of halos. For instance, with respect to their dark-matter-only (DMO) counterparts, hydrodynamic simulations tend to exhibit rounder halos \citep[e.g.][]{Butsky_2016, Chua_2019, Cataldi_2021, Chua_2021}, generally with less cuspy density profiles \citep[e.g.][]{Mashchenko_2008, Madau_2014, Oman_2015}. In particular, using the NIHAO \citep{Wang_2015} simulation, \cite{Maccio_2020} showed that the inclusion of AGN feedback makes the dark matter distribution in the inner regions of massive halos ($>3\times 10^{12} \Msun$) less cuspy. Similarly, results from the IllustrisTNG \citep{IllustrisTNG} simulation indicate that black hole kinetic winds have a major impact on the slope of the total density profile in early type galaxies, while stellar feedback plays a sub-dominant role in this respect \citep{Wang_2020}. On the other hand, \cite{Schaller_2015} showed that the presence of stars causes the density profiles of cluster-size halos from the \textsc{EAGLE} \citep{EAGLE_Schaye2015} simulation to be cuspier within 5\% of the virial radius if compared to its DMO counterpart. Note that the effect of baryons on the density profiles is also halo mass dependent \citep[e.g.][]{Cui_2014}. Thus, the exact effect of different baryon-driven physical mechanisms on the halo density profiles \citep[see][for comparisons between different simulations of the same galaxy cluster]{Cui_2016} is still an open research area.

The alterations to the density profile of halos induced by feedback naturally results in a variation of the enclosed mass. Within the NIHAO project, \cite{Tollet_2019} showed that galactic winds prevent gas accretion from cosmic filaments as far as six virial radii, reducing the mass of galaxies by a factor of $\sim2-4$. In the \simba\ simulation, AGN-driven jets are the dominant process that evacuates 80\% of baryons from halos by redshift $z=0$ \citep{Appleby_2021}; baryon particles can be moved out to as far as 15 Mpc \citep{Borrow_2020}. Other simulations such as IllustrisTNG, EAGLE and Magneticum \citep[e.g.][]{Dolag_2016} showed that more than half of the baryonic mass is displaced from Local Group-sized halos ($M_{500}>10^{12}-10^{13} \Msun$) due to feedback \citep{Lim_2021}. Furthermore, the impact of baryonic physics is manifest in the suppression of the number of subhalos \citep[e.g.][]{Sawala_2016, Zhu_2016, Elahi_2016, Chua_2017, Despali_2017} and in the break of the self-similarity of subhalo demographics \citep{Chua_2021}. Importantly, the action of feedback also impacts the thermal state of the gas in the circumgalactic medium \citep[CGM;][]{Suresh_2015, Turner_2017, Fielding_2020} and even intergalactic medium (IGM) \citep[e.g.][]{Christiansen_2020}, in a manner that can be constrained with observables such as absorption line statistics \citep{Rahmati_2013, Rahmati_2013b, Turner_2014, Rahmati_2015, Meiksin_2015, Keating_2016, Meiksin_2017, Viel_2017, Ravoux_2020, Sorini_2018, Sorini_2020, Appleby_2021}.

The impact of feedback has repercussions on the large-scale distribution of matter as well. For instance, results from the \textsc{Eagle}, \textsc{Illustris} and \textsc{IllustrisTNG} simulations show that the halo mass function is shifted to lower halo masses with the inclusion of baryons \citep{Beltz-Mohrmann_2021}. Results from the \simba\ simulation show that AGN jets significantly suppress the HI, H2 and stellar mass functions at the high-mass end \citep{Simba, Dave_2020}. Baryonic physics can also significantly affect cluster count cosmology \citep{Debackere_2020, Debackere_2021}, void statistics \citep{Paillas_2017}, as well as the power spectrum \citep{Hellwing_2016, Barreira_2019, van_Daalen_2020} and bispectrum \citep{Foreman_2020} of matter density fluctuations. An in-depth understanding of such effects is crucial for the interpretation of data from ongoing and forthcoming large-scale surveys (e.g., DESI, \citealt{DESI}; Euclid, \citealt{Euclid}; WEAVE, \citealt{WEAVE}), which often relies on large suites of numerical simulations \citep[e.g.][]{Martinelli_2021}.

There is thus a large body of literature on the effect of baryons on various aspects of galaxy formation. In this work, we will comprehensively explore the impact of baryons on halo density profiles and large-scale structure within the \simba\ cosmological hydrodynamic simulation suite including its dark-matter-only (DMO) counterpart. Taking advantage of several variants of the \textsc{Simba} simulation where different feedback modules are deactivated, we will also study the impact of feedback prescriptions. Compared with previous cosmological simulations, \simba\ is unique in its implementation of black hole accretion, which includes a torque-limited model for cold gas \citep{Hopkins_2011, Angles-Alcazar_2013, Angles-Alcazar_2015, Angles-Alcazar_2017a, Angles-Alcazar_2017b} alongside the usual Bondi accretion for hot gas \citep{Bondi_1952}. As the AGN feedback prescription is tied to the accretion of black holes, it is clearly of great interest to investigate how this novel model can affect the distribution and physical state of matter on a variety of scales. In our analysis, we will adopt a somewhat different view compared with past literature. Rather than focusing on the impact of the presence of baryons and of feedback processes on specific scales, we will aim at understanding how the effects of baryonic physics within halos and large scales are interconnected. In this way, we will be able to provide a unified picture for the multi-scale action of different feedback prescriptions. We will show that stellar winds and AGN-driven jets are the dominant physical drivers in shaping the distribution of matter in the Universe at redshift higher and lower than $z\sim 2$, respectively. This paper is a primarily theoretical study to gain insight on the physics regulating the distribution and physical state of matter in the Universe; we leave the comparison between specific observations and the predictions of \simba\ for future work.

We explain the main features of the \simba\ simulation in \S~\ref{sec:simulations}. We then discuss the effects of baryonic physics proceeding from large scales down to smaller ones. In \S~\ref{sec:IGM} we address the effect on the thermal state of the IGM, in \S~\ref{sec:mass_distrib} we consider the mass distribution of different baryonic phases across halos, and we investigate the density profiles within halos in \S~\ref{sec:profiles}. We present our conclusions in \S~\ref{sec:conclusions}. Throughout this manuscript, unless otherwise indicated, distances are expressed in proper units; comoving units are indicated with a `c' prefix (etc., $\cMpc$).

\section{Simulations}
\label{sec:simulations}

\begin{table*}
\begin{center}
\begin{tabular}{lcccccccc}
\hline
Simulation & Box size & Nr. of particles & DM particle mass & gas element mass & Stellar Feedback & AGN winds & Jets & X-ray heating \\
 & ($\hMpc$) & & ($\rm M_{\odot}$) & ($\rm M_{\odot}$) & & & & \\
\hline
Simba $100 \hMpc$ & 100  & $2\times 1024^3$ & $9.6 \times 10^7$ & $1.82 \times 10^7$ & \checkmark & \checkmark & \checkmark & \checkmark  \\
(fiducial-100) & & & & & & & & \\
Simba $50 \hMpc$ & 50  & $2\times 512^3$ & $9.6 \times 10^7$ & $1.82 \times 10^7$ & \checkmark & \checkmark & \checkmark & \checkmark  \\
(fiducial-50) & & & & & & & & \\
Simba $25 \hMpc$ & 25  & $2\times 256^3$ & $9.6 \times 10^7$ & $1.82 \times 10^7$ & \checkmark & \checkmark & \checkmark & \checkmark  \\
Simba High-res. & 25  & $2\times 512^3$ & $1.2 \times 10^7$ & $2.28 \times 10^6$ & \checkmark & \checkmark & \checkmark & \checkmark  \\
No-X-ray & 50  & $2\times 512^3$ & $9.6 \times 10^7$ & $1.82 \times 10^7$ & \checkmark & \checkmark & \checkmark  & \\
No-jet & 50  & $2\times 512^3$ & $9.6 \times 10^7$ & $1.82 \times 10^7$ &\checkmark & \checkmark & & \\
No-AGN & 50  & $2\times 512^3$ & $9.6 \times 10^7$ & $1.82 \times 10^7$ &\checkmark & & & \\
No-feedback & 50  & $2\times 512^3$ & $9.6 \times 10^7$ & $1.82 \times 10^7$ & & & & \\
Simba-Dark & 100 & $1024^3$ & $1.14 \times 10^8$ & --- & & & & \\
\hline
\end{tabular}
\caption{\simba\ runs used in this work.
}
\label{tab:runs}
\end{center}
\end{table*}

\simba\ is a suite of cosmological simulations based on the \texttt{Gizmo} hydrodynamic code. Specifically, gas particles are evolved following the meshless finite mass  (MFM) implementation of \texttt{Gizmo}, which allows for an accurate description of shocks and shear flows, without the need for any artificial viscosity \citep{Gizmo}. Thus, this feature guarantees faithful description of shocks and flows with high Mach number, as in the case of outflows and jets. Dark matter (DM) is represented through a set of collisionless Lagrangian particles solved via a tree-particle-mesh algorithm based on \texttt{Gadget}~\citep{Springel_2005}.

Radiative cooling and photoionisation heating are included through the \texttt{Grackle-3.1} library \citep{Smith_2017}, which accounts for metal cooling and the evolution of primordial elements out of equilibrium. \simba\ employs the \cite{HM12} uniform ionising background, modified to include self-shielding throughout the simulation run using the \cite{Rahmati_2013} prescription (A. Emerick, priv. comm.).

Star formation follows the same model adopted in the predecessor simulation \textsc{Mufasa} \citep{Mufasa}, which is based on a \cite{Schmidt_1959} law for $\rm H_2$, where the $\rm H_2$ is estimated from the local column density and metallicity as per the \cite{Krumholz_2011} prescription. Above a hydrogen number density $n_{\rm H}$ of $n_{\rm th}>0.13~{\rm cm}^{-3}$, we apply the minimal artificial pressurisation to the interstellar medium (ISM) that is necessary to resolve star-forming gas, such that the temperature of this gas has a lower limit of
\begin{equation}
    \log \left( \frac{T}{\rm K} \right) = 4 + \frac{1}{3}\log \frac{n_{\rm H}}{n_{\rm th}}.
\end{equation}
Gas with $n_{\rm H}>n_{\rm th}$ and with a temperature of at most 0.5 dex above this temperature floor is considered eligible to form stars, and we therefore define it as ISM. We stress that according to this definition not all ISM gas is actively forming stars, as it must also contain $\rm H_2$; at low metallicity, the density threshold to form $\rm H_2$ may be well above $n_{\rm th}$. The chemical enrichment model tracks eleven different elements  (H, He, C, N, O, Ne, Mg, Si, S, Ca, Fe) from Type Ia and II supernovae, and Asymptotic Giant Branch (AGB) stars \citep{Oppenheimer_2006}.

Star formation-driven galactic winds are described following kinetic decoupled ejection. Galactic outflows from massive stars are driven by a combination of Type II supernovae winds, radiation pressure and stellar winds, the aggregate effect of which is represented via a sub-grid prescription in which wind particles are ejected in the direction perpendicular to the plane identified by their velocity and acceleration vectors. The two main free parameters characterising such winds are the mass loading factor and the wind speed. The scaling of both these parameters with galaxy properties follows the rates predicted by the FIRE zoom-in simulations \citep{Muratov_2015, Angles-Alcazar_2017b}; see \citet{Simba} for full details.

The metallicity of the winds is metal-loaded to account for the Type II supernovae that generate the metals, by extracting metals from the surrounding ISM depending on the mass loading factor and Type II supernovae yields. A 30\% fraction of the ejected wind particles are heated to a temperature set by the difference between the supernova energy ($u_{\rm SN} = 5.165\times 10^{15}\ {\rm erg\ g}^{-1}$) and the kinetic energy; the remaining particles are ejected at $T\approx 10^3$K. Once wind particles are ejected, they are hydrodynamically decoupled to avoid numerical inaccuracies due to single gas elements with high Mach numbers relative to their surroundings. Furthermore, cooling is switched off too so that hot winds can deposit their thermal energy into the CGM. Outflowing wind particles are recoupled when at least one of the following conditions are true: the density of the particle is lower than that of the ISM and its velocity matches that of the surrounding particles; the particle density is below $0.01n_{\rm th}$; or the particle has been decoupled for a time of at least 2\% of the Hubble time at launch.

\simba\ includes black hole (BH) particles, which accrete following a dual model. Non-ISM gas with temperature $T>10^5 \K$ follows the Bondi accretion rate (`hot-accretion mode'). Otherwise, the gas within the BH kernel follows the `cold-accretion mode'. This is described with a torque-limited accretion model, driven by disk gravitational instabilities arising from galactic scales down to the accretion disk around the central BH (\citealt{Hopkins_2011}; see also \citealt{Angles-Alcazar_2013, Angles-Alcazar_2015, Angles-Alcazar_2017a}). There are three different ways in which the AGN feedback is implemented, depending on the mass and accretion rate of the black hole. Fast-accreting BHs ($>0.2$ times the Eddington accretion rate) eject \textit{radiative winds}, modelled as purely bipolar outflows, with a direction parallel to the angular momentum of the BH. The radiative wind velocity scales as:
\begin{equation}
    \frac{v_{\rm AGN \, w}}{\rm km \, s^{-1}}=500+\frac{500}{3} \left(\log \frac{M_{\rm BH}}{\Msun}-6 \right) \, .
\end{equation}
The winds are then kinetically coupled to the surrounding gas particles. Consistent with observations of ionised gas outflows, implying electron temperatures of order $10^4 \K$, radiative winds do not directly affect gas temperature, which is still set by the aforementioned ISM pressurisation model. In BHs with mass $>10^{7.5}  \, \rm M_{\odot}$, as the accretion rate falls below the $0.2$ Eddington threshold, AGN feedback transitions to the \textit{jets} mode feedback (in line with indications from observations such as \citealt{Barisic_2017}). AGN jets are still ejected in the form of purely bipolar outflows, but can reach much higher velocities, as dictated by the following equation:
\begin{equation}
    \frac{v_{\rm AGN \, jet}}{\rm km \, s^{-1}}=\frac{v_{\rm AGN \, w}}{\rm km \, s^{-1}} + 7000 \log \left( \frac{0.2}{f_{\rm Edd}} \right) \, .
\end{equation}
Thus, the AGN jets mode will becomes progressively more prominent as the accretion rate decreases. However, the velocity boost is capped at $7000 \, \rm km/s$ when the Eddington ratio reaches  $f_{\rm Edd}\leq 0.02$. Finally, BHs with active AGN jets can also exert \textit{X-ray heating} feedback if the gas fraction of the host galaxy is lower than $0.2$. X-ray heating affects only the gas particles within the BH kernel, and is proportional to the inverse square of the distance of the gas particle from the BH. Within the kernel, the temperature of the non-ISM gas is increased based on the local heating flux, while for ISM gas half of the X-ray energy is added as heat, and the other half is converted into kinetic energy by imparting a radial outwards kick to the gas particles. In this way, low-resolution ISM is prevented from the quick cooling that would be induced by the ISM pressurisation model \citep{Mufasa}.

\subsection{Runs}

The results presented in this work are based on six runs of the \simba\ suite of hydrodynamic simulations. The flagship run (fiducial-100) is a $100 \hMpc$ box with $1024^3$ DM particles and as many gas elements, with mass resolutions of $9.6 \times 10^7 \, M_{\odot}$ and $1.82 \times 10^7 \, M_{\odot}$, respectively. This run contains all physical prescriptions described earlier in this section. The simulation follows a \LCDM cosmological model consistent with \cite{Planck_2016} cosmological parameters ($\Omega_{\mathrm{m}}=0.3$, $\Omega_{\Lambda}=1-\Omega_{\mathrm{m}}=0.7$,
$\Omega_{\mathrm{b}}=0.048$, $h=0.68$, $\sigma_8=0.82$, $n_s=0.97$, with the usual definitions of the parameters). We then consider a smaller version of the flagship run (fiducial-50), with a box size of $50 \hMpc$ and the same resolution. Additionally, we run four more variants of the fiducial-50 run, where different feedback modules are progressively deactivated, as summarised in Table~\ref{tab:runs}. These runs start from the same initial conditions as in the fiducial-50 simulation. We could not explore the various AGN feedback prescriptions in a suite of $100 \hMpc$ \simba\ simulations with $2\times 1024^3$ particles as we did for the $50 \hMpc$ runs because of the computational resources available.  We further consider a DMO version of the \simba\ flagship run (\simba-Dark), with the same box size, number of DM particles and initial conditions, with the obvious exception of no gas elements. Finally, in order to perform convergence tests, we considered two smaller \simba\ runs, with a box size of $25 \hMpc$ and $512^3$ and $256^3$ particles, respectively.

In all runs, halos are identified on the fly via a 3D friends-of-friends algorithm embedded in \texttt{Gizmo}, based on the code by V. Springel in \texttt{Gadget-3}. A linking length equal to 0.2 times the mean inter-particle separation is adopted. We run the \texttt{yt}-based package \textsc{Caesar} \footnote{\url{https://caesar.readthedocs.io/en/latest/}} in post processing in order to cross-match galaxies and halos. \textsc{Caesar} also produces a catalogue with many relevant pre-computed properties of galaxies and halos. Many results of this work are obtained by analysing such catalogues.

\section{Mass distribution outside halos}
\label{sec:IGM}

A good starting point to understand the large-scale distribution of baryons is to compute the mass fraction of baryons that are locked in the IGM as a function of redshift. To do this, we consider all snapshots of the fiducial-100 run corresponding to the redshift range $0<z<6$. For every snapshot, we identified all gas particles that do not belong to any halo. In this context, a particle is considered to be part of a halo if its distance from the minimum of the gravitational potential does not exceed the virial radius $r_{200}$, i.e. the radius of the sphere containing an average density equal to 200 times the critical density of the Universe. The mass fraction of baryons locked in the IGM, $f_{\rm IGM}$, is then simply given by the total mass of all gas elements\footnote{We verified that the number of star and BH particles lying outside halos is negligibly small at all redshift considered, as expected. Therefore, equating the `baryon mass fraction outside halos' and the `gas mass fraction outside halos' is justified.} outside halos, divided by the total baryon mass within the simulation box at the redshift of interest.

\begin{figure}
    \centering
    \includegraphics[width=\columnwidth]{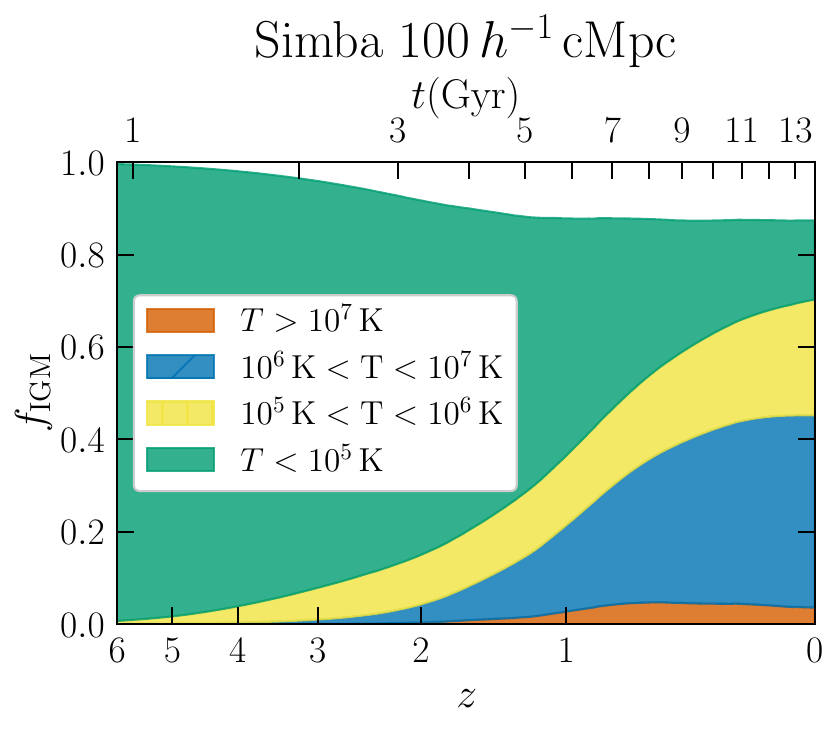}
    \caption{Redshift evolution of the mass fraction of gas in the IGM, as defined in the main text, in the \simba\ $ 100 \hMpc$ run. The orange, blue, dark yellow and orange shaded areas refer to gas within different temperature ranges, as reported in the legend of the figure. At $z=0$, 84\% of the baryon mass in the Universe is locked in the IGM. Note that the unshaded (white) region therefore represents the baryonic mass fraction within halos.}
    \label{fig:fIGM}
\end{figure}

We show the results of this calculation in Figure~\ref{fig:fIGM}. The lower horizontal axis shows the redshift, while the upper axis the corresponding cosmic time. We also highlight the contribution to $f_{\rm IGM}$ from gas at different temperatures: the green shaded area refers to the cool IGM ($T<10^5 \K$), and the other colours represent different phases of the warm-hot intergalactic medium (WHIM). Specifically, the blue and dark yellow areas refer to gas with temperature in the intervals $10^5 \K - 10^6 \K$ and $10^6 \K - 10^7 \K$, respectively, and the orange area to hot gas with $T>10^7 \K$.

\begin{figure*}
    \centering
    \includegraphics[width=\textwidth]{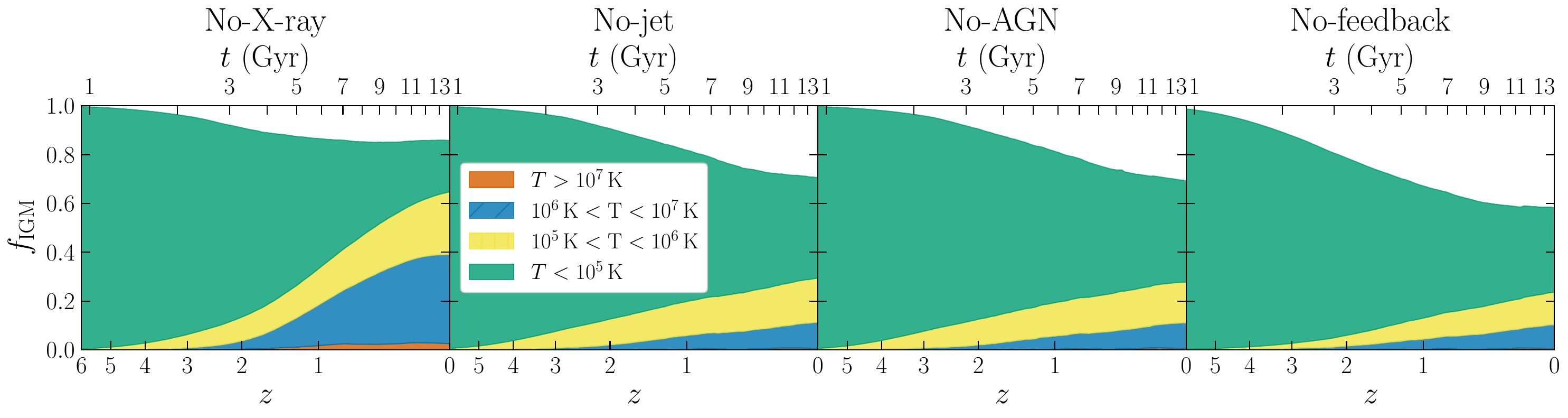}
    \caption{Same as in Figure~\ref{fig:fIGM}, but for the different \simba\ feedback variants. Each panel refers to a different run, as indicated in the upper part of the figure. AGN-driven jets are crucial for transferring baryons outside halos and heating the IGM.}
    \label{fig:fIGM_feedback}
\end{figure*}

We note that at $z=6$ almost the entirety of the baryonic mass resides in the IGM, and then it gradually decays, reaching $\sim 87\%$ at $z\approx 1.2$, as a consequence of gas accretion onto halos. The value of $f_{\rm IGM}$ is essentially unchanged after $z=1.2$, and is in excellent agreement with observational constraints indicating that the amount of baryons within halos at $z=0$ is approximately 17\% \citep{Shull_2012}. Our findings are also consistent with the results obtained by \cite{Cui_2019} by separating baryons into different environments, with the knots of the cosmic web (corresponding to halos) containing about 10\% of the total baryonic mass at $z\lesssim 1$. Clearly, at high redshift the majority of the IGM is still in the cold phase. As we proceed to lower redshift, star formation progressively increases, hence activating feedback processes that act as a source of heating. Additionally, BH growth triggers AGN feedback processes that contribute to the heating and expulsion of gas from haloes. Therefore, there is progressively a larger amount of gas in the WHIM phase at late times. To understand to what extent different physical processes cause the observed evolution of $f_{\rm IGM}$ in the simulation, we need to repeat our computations with the \simba\ variants that follow alternative feedback prescriptions. Even though these runs involve a smaller volume ($50 \hMpc$), a direct comparison with the results of the \simba\ $100 \hMpc$ run is still possible, as we verified that Figure~\ref{fig:fIGM} looks almost exactly the same for \simba\ $50 \hMpc$.

We now report the results for $f_{\rm IGM}$ in the different \simba\ runs in Figure~\ref{fig:fIGM_feedback}, where we adopt the same colour-coding for the contribution of the various gas phases as in Figure~\ref{fig:fIGM}. The left-most panel shows the results for the No-X-ray run, which are hardly distinguishable from those found for the \simba\ $100  \hMpc$ run. Although the amount of gas in the WHIM phases is about $6\%$ lower than in the fiducial-100 run over the entire redshift range, the predicted evolution of $f_{\rm IGM}$ is very similar. This suggests that the presence of X-ray heating does not have any appreciable large-scale effect on the IGM. This result is physically sensible, as by construction the X-ray mode of AGN feedback acts only within the kernel of the central BH, and as such is not expected to cause any major impact on material outside halos.

On the contrary, additionally switching off AGN-driven jets drastically changes the evolution of $f_{\rm IGM}$. As we can see in the second panel from the left in Figure~\ref{fig:fIGM_feedback}, in the No-jet run the IGM is dominated by the cool phase at all times, and the WHIM contains virtually no gas with $T>10^7 \K$. This is in stark contrast with the results obtained with the \simba\ $100 \hMpc$ run, especially for $z\lesssim 2$. This is not surprising, as it is only after $z\approx 2$ that most BHs have grown enough such that their accretion rate drops below the threshold necessary to activate the AGN jet feedback module \citep{Christiansen_2020}. In the No-jet run, at $z=0$ the gas in the temperature ranges $T<10^5 \K$, $10^5 \K-10^6 \K$ and $10^6 \K - 10^7 \K$ contributes by 41.2\%, 18.1\% and 10.7\%, respectively. The corresponding values in the fiducial-100 run are 17.1\%, 25.1\% and 41.6\%, not to mention that there is a non-negligible fraction of hot $T>10^7 \K$ gas (3.6\%). Therefore, at $z=0$, $f_{\rm IGM}= 70\%$ in the No-jet run and $f_{\rm IGM}= 87.4\%$ in the fiducial-100 run. This means that AGN-driven jets are crucial in both heating gas, and transferring hot gas from halos to the IGM. 

Our results are consistent with the findings by \cite{Christiansen_2020}, who computed the mass fraction of different baryonic phases within the in the \simba\ $50 \hMpc$ and No-jet runs at $z=0$. However, they did not classify gas elements as `IGM' based on their proximity to halos. Instead, gas elements above a certain overdensity threshold $\Delta_{\rm th}$ were considered to be `halo particles'. Such threshold was determined following the estimate by \cite{Dave_2010} of the typical overdensity relative to $\Omega_{\rm m}$ at the virial radius of halos at redshift $z$ (in turn based on \citealt{Kitayama_1996}). For the cosmological model embedded in \simba, $\Delta_{\rm th} \approx 110$ at $z=0$. \cite{Christiansen_2020} found that the mass fraction of gas and stars above $\Delta_{\rm th} \approx 105$ was 13.1\% and 32.5\% for the \simba\ $50 \hMpc$ and No-jet runs, respectively. This is in excellent agreement with our results, which are 16.2\% and 30\%, respectively. \cite{Christiansen_2020} further classify the baryon particles with $\Delta_{\rm th}<105$ according to their temperature: if $T>10^5 \K$, they are considered to be in the `WHIM' phase, otherwise in the `diffuse' phase. Following this criterion, \cite{Christiansen_2020} found that the mass fraction in the WHIM and diffuse phases is 70.5\% (28.7\%) and 16.4\% (38.8\%) for the \simba\ $50 \hMpc$ (No-jet) run. Once again, these values are in good accord with our results: we find that in the \simba\ $100 \hMpc$ (No-jet) run 70.3\% (29.4\%) and 17.1\% (41.2\%) of the baryon mass outwith halos is in the $T>10^5 \K$ and $T<10^5 \K$ phase, respectively. Thus, our approach effectively validates the prescription adopted by \cite{Christiansen_2020} to distinguish between `halo' and `IGM' particles. The results discussed in this section also represents an extension of \cite{Christiansen_2020} work to earlier redshift, and to different feedback variants. Indeed, we will now investigate how the evolution of $f_{\rm IGM}$ changes if we turn off AGN feedback altogether.

The No-AGN run results are shown in the third panel from the left in Figure~\ref{fig:fIGM_feedback}. We can immediately see that there is no significant difference with respect to the No-jet run. Quantitatively, the split of the baryonic mass among the different IGM phases differs by at most 2\% over the full redshift range. This result clearly shows that the impact of radiative AGN winds on the thermal state of the IGM is sub-dominant with respect to that of AGN jets. It also strongly suggests that the kinematic impact of radiative winds is confined within the virial radius of halos. However, in order to validate this hypothesis it is necessary to investigate the halo density profiles of different baryonic phases. We will do this in \S~\ref{sec:profiles}.

We finally show the evolution of $f_{\rm IGM}$ in the No-feedback run in the right-most panel in Figure~\ref{fig:fIGM_feedback}, where additionally star formation winds have been turned off. The trend of the baryon mass fraction for gas with $T>10^5 \K$ is similar as in the No-AGN and No-jet runs, except that the total amount of gas is lower. Specifically, the mass fraction of gas particles with temperatures in the ranges $10^5\K-10^6\K$ and $10^6\K-10^7\K$ at $z=0$ is 12\% and 10\%, respectively, in the No-feedback run. The amount of cool gas ($T<10^5\K$) is 18\%, thus taking the total baryon mass fraction in the IGM at $z=0$ to 60\%. This is about 10\% lower than in the No-AGN case. Furthermore, the slope at which $f_{\rm IGM}$ decays is steeper than in the No-AGN run. Whereas in the No-AGN run we need to wait $z\approx 2$ for $f_{\rm IGM}$ to fall below 90\%, such threshold is crossed already at $z\approx 3 $ in the No-feedback run. This reflects the fact that stellar feedback is efficient already at high redshift. The action of supernovae-driven winds thus contribute to the gas heating and depletion of halos, more significantly at high redshift. By comparing to the No-AGN run, the SN feedback seems even more powerful than the thermal AGN feedback in setting the thermal state of the IGM; at $z\approx 2-3$, AGN jets gradually overtake stellar feedback as the most effective heating source for the IGM. We thus expect that observational probes that are sensitive to state of  the low-redshift IGM, such as fast radio bursts, will provide useful constraints for feedback models \citep[see, e.g.,][]{Lee_2022}.

Having thoroughly explored how different baryonic physics shapes the IGM phases, it is now natural to ask how baryon-driven physical processes impact the mass distributions in various components as a function of halo mass. This will be the subject of the next section.

\section{Mass distribution across halos}
\label{sec:mass_distrib}

In this section, we will investigate the distribution of matter across halos of different mass. To begin with, we will explore how the presence of baryons and the processes that they trigger affect large-scale structure statistics such as the halo and baryon mass functions (\S~\ref{sec:mass_funcs}). We will then consider individual halos in the fiducial-50 run, and their counterparts in the different feedback variants, to determine how their stellar, baryon and total mass change depending on the run (\S~\ref{sec:mass_content}). Finally, we will investigate how much different phases of baryonic matter contribute to the total baryon mass of halos at different redshifts, and for different feedback prescriptions (\S~\ref{sec:omega}).

\subsection{Mass functions}
\label{sec:mass_funcs}

The halo mass function (HMF) is one of the most widely used large-scale structure statistics. For a fixed set of cosmological parameters, its redshift evolution immediately provides information on the clustering of matter over cosmic time. Thus, it has been the subject of several theoretical studies. Early analytical works provided a physically motivated shape for the HMF \citep[e.g.][]{PS74, ST99} and subsequent work proposed analytical fits to the HMF obtained in simulations \citep[e.g.][]{Jenkins_2001, Warren_2006}. Such fitting formulae hinted towards the universality of the HMF; this was explicitly verified through comparison with numerical simulations \citep{Despali_2016}. However, this remarkable result has some limitations, as the HMF is sensitive to the exact definition of halos and on the halo finding algorithm used \citep[e.g.][]{Lukic_2007, Tinker_2008,  Manera_2010, Watson_2013}. Other works focused on the impact of baryons on the halo mass function. The HMF appears to be a robust statistic, with discrepancies between hydrodynamic and N-body runs being within 20\% \citep[e.g.][]{Cui_2012, Castro_2021}.

\begin{figure*}
    \centering
    \includegraphics[width=\textwidth]{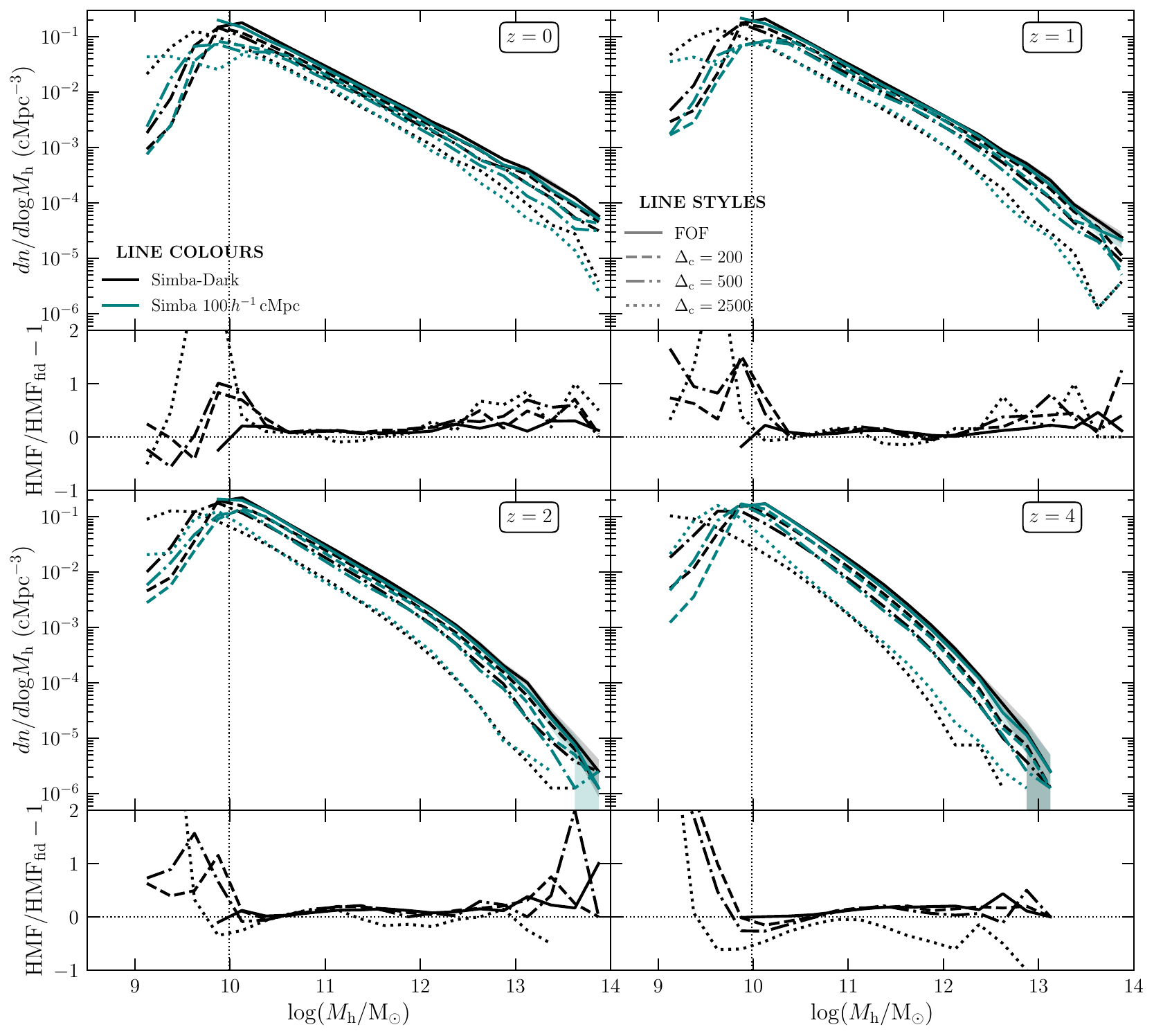}
    \caption{Halo mass function in the \simba\ $100 \hMpc$ and \simba-Dark simulations (teal and black lines, respectively), for different definitions of the halo boundaries (represented with different line styles), as explained in the main text. The larger panels refer to different redshifts; the redshift is indicated inside each panel. The shaded areas represent the scatter due to cosmic variance (see main text for details). This is indicated only for $z=0$, not to overcrowd the plot. The vertical dotted line corresponds to halos with at least 100 DM particles. The smaller panels beneath each larger panel show the relative difference of the halo mass functions in the \simba-Dark run with respect to the \simba\ $100 \hMpc$ run, for the same definition of the halo boundaries. Especially at $z\geq2$, the relative differences are generally larger when halos are defined such that their density is 2500 times the critical density of the Universe.}
    \label{fig:hmf_dm}
\end{figure*}

With the perfection of observational techniques to estimate the baryon content of halos came the possibility of investigating another interesting large-scale structure statistic: the baryon mass function (BMF). Analogously to the HMF, this quantity encodes information on the number density of halos as a function of their baryon mass. Several observations provide us with data up to $z=3$ \citep{Read_2005, Eckert_2016, Pan_2019}. However, in this work we are mainly interested in the more theoretical aspects of the subject, and rather than the observable \textit{galaxy} baryon mass function, we will consider the \textit{halo} baryon mass function, i.e. the number density of halos per baryon mass bin. We wish to investigate how both the HMF and BMF are impacted by baryons in \simba.

As a starting point, we want to understand how the HMF is affected by the mere presence of baryons as opposed to a hypothetical DM-only universe. Thus, in Figure~\ref{fig:hmf_dm} we compare the HMF obtained with the \simba\ $100 \hMpc$ (teal lines) and \simba-Dark runs (black lines), at different redshifts (as indicated inside the main panels), and for different definitions of the boundaries of halos. Specifically, we show the results obtained by defining the halo mass as the mass of all particles within the FOF boundaries of halos (solid lines), or within a sphere centred at the minimum of the gravitational potential and with a radius chosen such that the enclosed total mass density equals $\Delta_{\rm c} = 200$, $\Delta_{\rm c} =500$ and $\Delta_{\rm c} =2500$ times the critical density of the Universe (dashed, dot dashed, and dotted lines, respectively). 

The bigger panels show the number density of halos per logarithmic halo mass ($M_{\rm h}$) bin. The vertical dotted line represents the mass of 100 DM particles in the halos considered ($\sim 10^{10} \Msun$). We do not consider halos with less than 100 DM particles to be well resolved. Note that the resolved halos with $\Delta_{\rm c} =2500$ definition should be slightly shifted toward lower halo mass as the enclosed mass is lower than the mass obtained with the FOF algorithm for the same halo. Therefore, the drop of the HMF for $M < 10^{10} \Msun$ is likely to reflect a resolution issue, and is thus spurious. The shaded areas around the HMFs depicted in the bigger panels represent the scatter of the HMF due to cosmic variance. This quantity is calculated by removing all halos within one octant of $50 \hMpc$ from the simulation at a time, then computing the HMF in the remaining seven octants, and finally taking the standard deviation of the eight estimates. The spread due to cosmic variance is evident only at the high-mass end at all redshifts, thus reflecting the rarity and spatial inhomogeneity of large halos. 

In the smaller panels below each of the larger panels, we show the relative difference between the HMF of the DMO run with respect to its counterpart in the fiducial \simba\ run, for all definitions of the halo boundaries considered. The horizontal dotted line simply marks the zero level, i.e. the level of perfect agreement. We notice that at all redshifts the HMF obtained with the \simba-Dark run matches the corresponding HMF in the fiducial-100 run within 20\% in the halo mass range $10^{10} \Msun < M_{\odot} < 10^{12} \Msun$, except for the case where $\Delta_{\rm c} = 2500$ at $z=2$ and $z=4$. It is expected that the presence of baryons has a more visible effect for the choice $\Delta_{\rm c} = 2500$: with this definition, the virial radius is smaller, and several works showed that the impact of baryons on the density profile within halos is more pronounced in the core \citep{Mashchenko_2008, Madau_2014, Pontzen_2012, Oman_2015, Onorbe_2015, Agertz_2016}. As we will later show, baryons appear to be indeed more concentrated in the centre of halos at higher redshift (Figure~\ref{fig:Omprof}). Interestingly, for all other definitions of the halo boundaries, the \simba-Dark results match the HMF of the fiducial-100 run with essentially the same precision. 

We notice that the relative difference between \simba-Dark and \simba\ $100 \hMpc$ is larger for $M_{\rm h}<10^{10} \Msun$, and at the high mass end ($M_{\rm h}>10^{12}-10^{13}\Msun$, depending on redshift). As mentioned above, low-mass halos are not well resolved, therefore the HMF is not reliable for $M_{\rm h }<10^{10} \Msun$. Regarding the high-mass end, the larger relative differences can be ascribed to the lower number of massive halos (see the Appendix~\ref{sec:convergence} for further convergence tests on the HMF). At $z<1$, there are typically $30-40$ and $10-20$ haloes in the largest mass bin ($M>10^{13.75} \Msun$) for $\Delta_{\rm c}=200$ and $\Delta_{\rm c}=500$, respectively, while for $\Delta_{\rm c}=2500$ the number of such massive haloes drops to $1-3$. At higher redshift, halos become scarcer at lower masses; the number of haloes is of order unity in all $M\gtrsim 10^{12.6} \Msun$ bins regardless of the definition of the halo boundaries.

A reduction in the number of haloes at high masses is to be expected due to the finite size of the box. It was shown by \cite{Sirko_2005} that because of their finite volume, N-body simulations tend to underestimate the variance of density fluctuations within spheres of a given size. Thus, clustering is suppressed and massive haloes are underrepresented. However, the number of haloes with mass $M\sim 10^{13} \Msun$ is enhanced by the cutoff of large-scale modes \citep{Power_2006}. \cite{Reed_2007} developed a technique to correct for the aforementioned effects, and showed that the abundance of rare haloes forming in an overdensity corresponding to a $5\sigma$ statistical fluctuation found in N-body simulations can be reduced by about 50\% from the predictions of a Sheth \& Tormen HMF. However, even if we counteracted this effect by increasing the number of massive haloes by 50\%, the trend of the evolution of the HMF at the high-mass end would remain almost unaffected. Also, the reduction in the number of haloes should affect in the same way the HMFs calculated at the same snapshot, for all definitions of the halo boundaries considered. Thus, we expect that our conclusions from the relative comparison across different choices of $\Delta_{\rm c}$ would be largely unaffected by correcting for the finite size of the box. We also verified that combining all haloes in the same bin for $M \gtrsim10^{13}\Msun$ would not qualitatively change our conclusions.

\begin{figure}
	\includegraphics[width=\columnwidth]{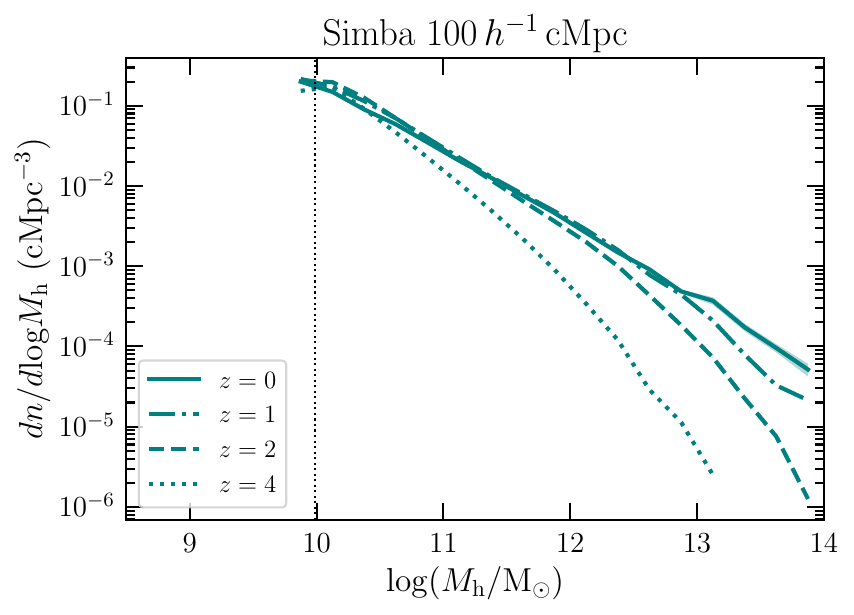}
	\includegraphics[width=\columnwidth]{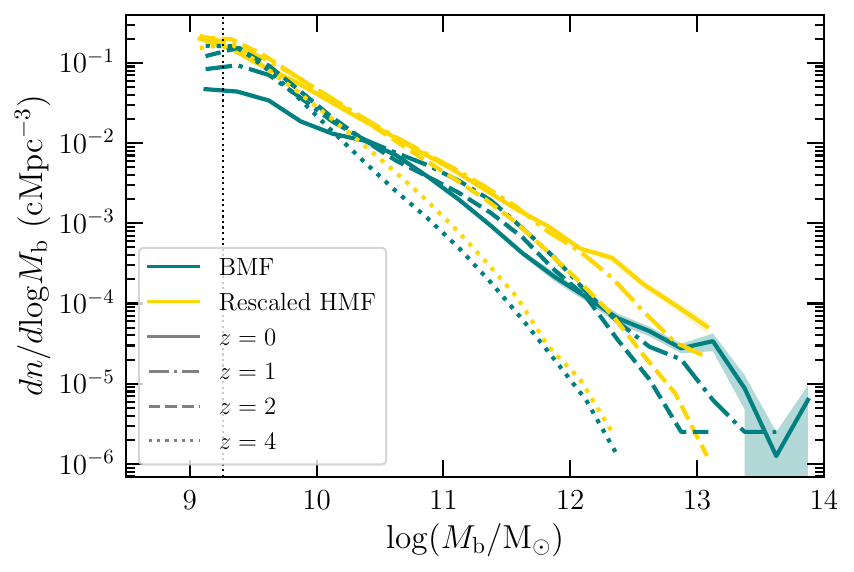}
    \caption{From top to bottom, halo mass function and baryon mass function of all halos in the \simba\ $100 \hMpc$ run, at different redshift. In all panels, the solid, dot-dashed, dashed and dotted lines refer to the snapshots at $z=0$, $z=1$, $z=2$ and $z=4$, respectively. The teal shaded area around the $z=0$ lines shows the scatter in the mass functions due to cosmic variance, calculated as explained in the main text. The yellow curves are the halo mass functions from the top panel, multiplied by $f_{\rm b}$ along the $x$-axis. The vertical dotted lines in the upper and lower panel correspond to a mass of 100 DM particles and 100 gas elements, respectively, and as such serve as a guide above which halos can be considered to be well resolved. The baryonic mass function is similar to the rescaled halo mass function at high-$z$, but by $z=0$ it lies substantially below.}
    \label{fig:mfuncs}
\end{figure}

To summarise, for the \simba\ simulation there appears to be no significant impact of the presence of baryons on the resulting HMF, except for the case where the boundaries of halos are defined following the $\Delta_{\rm c}=2500$ convention. For all other definitions, the \simba-Dark run matches the \simba\ fiducial-100 run within the same level of precision. Hereafter, unless otherwise indicated, we will follow the $\Delta_{\rm c}=200$ definition for the halo boundaries, which  will thus constitute our preferred choice. To highlight the redshift evolution of the HMF in the fiducial-100 run for this definition of the halo boundaries, we again show it in the upper panel of Figure~\ref{fig:mfuncs}, with the same colour coding adopted in Figure~\ref{fig:hmf_dm}. The line styles now refer to different redshifts, as indicated in the legend of the plot. At higher redshift, the HMF exhibits a cutoff at smaller halo mass. Conversely, at lower redshift the amount of high-mass halos increases. This reflects the evolution of clustering in the Universe: as time goes by, it will become more likely to form higher mass halos until the growth of structure eventually freezes out as $\Lambda$ starts dominating. We cannot observe such freeze-out by evolving the simulation until $z=0$.

We can now compute the BMF predicted by \simba, of course for the fiducial-100 run only, as the \simba-Dark run does not contain baryons. We plot the BMF at different redshifts with the teal lines in the lower panel of Figure~\ref{fig:mfuncs}, where the line styles have the same meaning as in the upper panel. The vertical dotted line now corresponds to 100 times the mass of a single gas element. Halos with baryon mass below this threshold are considered to not be sufficiently resolved. 

The redshift evolution of the BMF is not as straightforward as that of the HMF. Whereas the number density of halos with high baryonic mass increases at lower redshift due to halo accretion, the BMF in the baryonic mass range $10^{10.5}\Msun \lesssim M_{\rm b} \lesssim 10^{12} \Msun$ starts decreasing after $z=1$. As we will show later, the higher efficiency of AGN jets after $z<2$ is responsible for evacuating gas from halos and suppressing star formation in halos with masses consistent with the aforementioned range of $M_{\rm b}$. We also note that the spread of the BMF due to cosmic variance at $z=0$ is larger than in the  case of the HMF. This is because of the rarity of halos with $M_{\rm b}>10^{13} \Msun$.

To highlight the role of baryons in shaping the evolution of the BMF, we also compute the `rescaled HMF'. In other words, we (wrongly) assume that the baryon mass within all halos can be obtained by simply re-scaling the total halo mass by the cosmic baryon mass fraction, $f_{\rm b}=\Omega_{\rm b} / \Omega_{\rm m}$. We then plot again the HMF, but this time as a function of $f_{\rm b} M_{\rm h}$ rather than $M_{\rm h}$. This `rescaled HMF' is plotted with dark yellow lines in the lower panel of Figure~\ref{fig:mfuncs}. The line style represents redshift, following the same convention as for the BMF. 

\begin{figure*}
	\includegraphics[width=\textwidth]{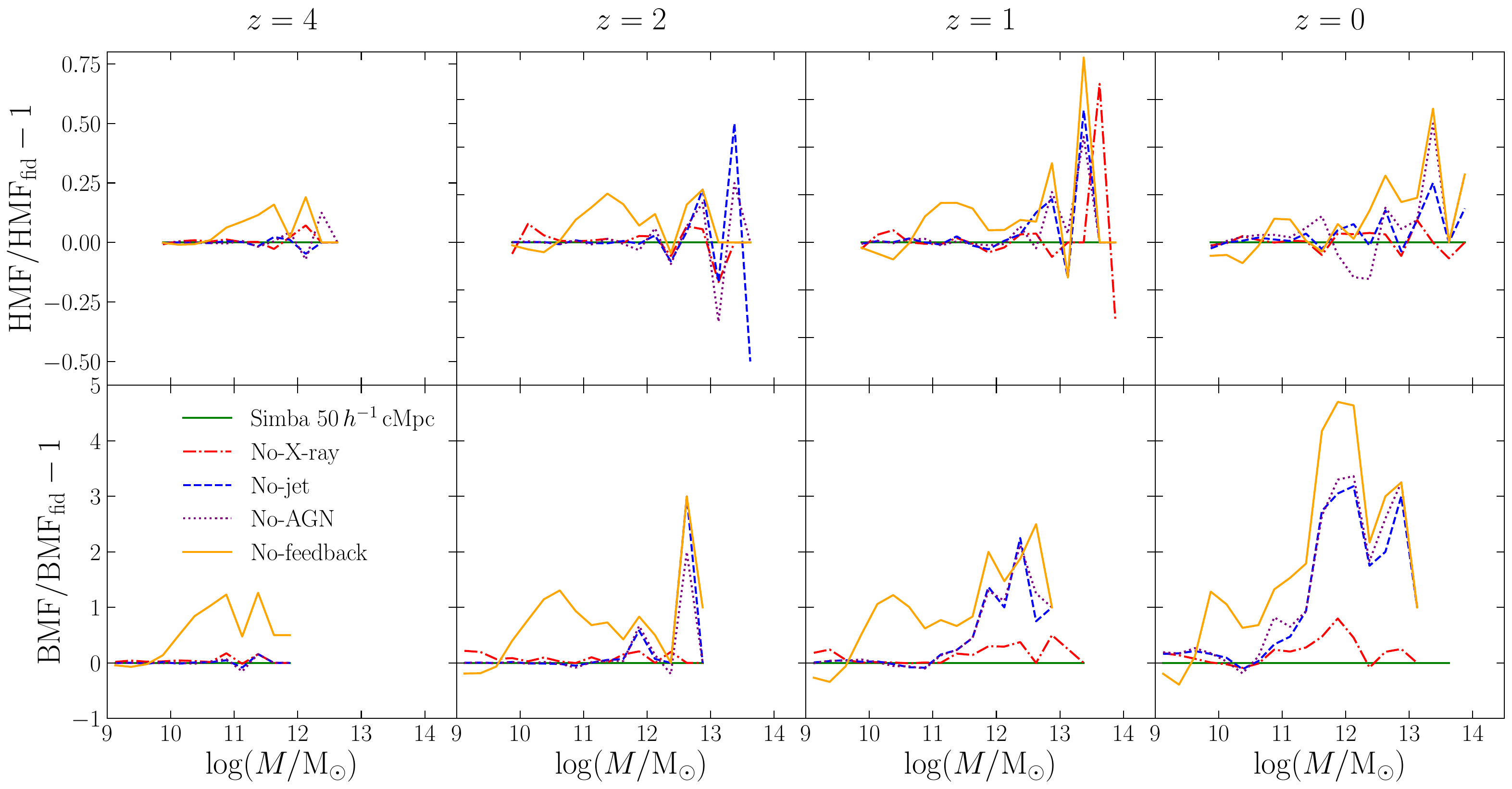}
    \caption{Relative differences in the mass functions across the various $50 \hMpc$ runs, for different redshifts. The panels in the top and bottom rows refer to the halo and baryon mass function, respectively. Thus, the masses on the $x$-axis refer to the total mass and baryonic mass in halos, respectively. All panels in the same column show the results at the redshift reported at the top. In all panels, the dot-dashed red, dashed blue, dotted purple and solid orange lines refer to the No-X-ray, No-jet, No-AGN and No-feedback runs, respectively. Jets are key to suppress the baryon and gas mass functions at the high-mass end.}
    \label{fig:mfuncs_diff}
\end{figure*}

At $z=4$, the rescaled HMF closely resembles the actual BMF, suggesting that the accretion of baryons tends to mostly follow halo growth, and the approximation $M_{\rm b} \approx f_{\rm b} M_{\rm h}$ is fairly justified. However, at $z=2$ we can already observe a suppression (albeit modest) of the BMF at $M_{\rm b} \sim 10^{11} \Msun$ with respect to the rescaled HMF. At even lower redshifts, the rescaled HMF overestimates the BMF. At the high-mass end, this probably mainly due to the AGN jet feedback prescription, which can more efficiently remove gas from halos. At the low-mass end, stellar feedback is likely to be the main driver of the suppression. To verify whether this is actually the case, we will now analyse the HMF and BMF in all the \simba\ variants.

In Figure~\ref{fig:mfuncs_diff} we show the relative difference of the HMF and BMF obtained in different runs with respect to the fiducial-50 run, with the same initial conditions. We chose this simulation as a reference because it is run with the same initial conditions and box size as the runs with alternative feedback prescriptions. We verified the BMF and HMF obtained with the \simba\ $50 \hMpc$ run generally match those given by the fiducial-100 run within $\sim 20\%$ (see the Appendix~\ref{sec:convergence} for details), hence they are well converged volume wise.

In the first row of panels in Figure~\ref{fig:mfuncs_diff}, we analyse the relative differences in the HMF; the red, blue, purple and orange lines refer to the No-X-ray, No-jet, No-AGN and No-feedback runs, respectively. Every panel corresponds to a different redshift, as specified in the upper part of the figure. The second row of panels in Figure~\ref{fig:mfuncs_diff} shows the relative differences in the BMF, following the  same colour coding as in the upper panels. The horizontal axis represents the total halo mass for the upper panels, and the baryonic halo mass for the lower panels.

The runs with at least stellar feedback yield very similar HMFs with respect to the fiducial-50 run. The scatter around the HMF of the \simba\ $50 \hMpc$ run grows with decreasing redshift, reaching at most $\sim 20\%$ for $M_{\rm h} \lesssim 10^{13} \Msun$. The relative difference can grow up to $60\%$ at higher masses, but the estimate of the HMF is less precise in this regime due to the larger cosmic variance and and the lower number of haloes in the higher-mass bins. The No-feedback run exhibits a larger relative difference already since $z=4$. The maximum relative difference in this run grows from $\sim 17\%$ at $z=4$ up to $\sim 40\%$ at $z=0$. Rather than oscillating around the HMF of the \simba\ $50 \hMpc$ simulation, the No-feedback run tends to systematically overestimate the HMF for $M>10^{10} \Msun$. At lower masses, the HMF appears to be underestimated, however halos are not well resolved in this regime.

Whereas the HMF proves to be a rather robust quantity across different feedback runs, the BMF is much more sensitive to baryonic physics. At $z=4$, the  BMF of the No-feedback run is systematically larger than in the \simba\ $50\hMpc$ simulation, by about a factor of $\sim 2$ for $M>10^{10.5} \Msun$. On the other hand, all other runs match the \simba\ $50 \hMpc$ simulation within $\sim  10\%$. This suggests that at high redshift stellar feedback is sufficient for diminishing the overall amount of gas accretion in halos. In the absence of any feedback process, there is nothing preventing halos from accreting baryons except for their own pressure and shock heating, hence increasing the number density of halos at any $M_{\rm b}$. This scenario remains substantively unchanged at $z=2$, as the largest relative differences between the \simba\ $50 \hMpc$ simulation and the runs that include at least stellar feedback occur at the high-mass end, where the cosmic variance on the BMF is larger, and therefore its estimate is less precise. 

At $z<2$, the No-feedback run keeps overestimating the BMF. At $z=0$, the number density of halos with $M_{\rm b} \approx 10^{12} \Msun$ is $\sim 6$ times larger than in the fiducial-50 run, while the No-AGN and No-jet runs follow almost the same trend, albeit with smaller changes. In both these runs, there is an excess of halos with $M_{\rm b}\gtrsim 10^{11} \Msun$ with respect to the \simba\ $50 \hMpc$ run. This excess results in the BMF being up to 3 and 4 times larger than in the \simba\ $50 \hMpc$ simulation at $z=1$ and $z=0$, respectively. In contrast, switching off X-ray heating has a much more moderate impact of $<50\%$ across all redshifts considered. Thus, AGN jets are crucial in shaping the BMF at the high mass end at low redshift, while stellar feedback is the main physical driver in the suppression of the BMF at high redshift. The results shown in Figure~\ref{fig:mfuncs_diff} hence confirm our aforementioned expectations.

We conclude by pointing out that the HMF and BMF are global statistics. As such, Figure~\ref{fig:mfuncs_diff} tells us how feedback affects the total and baryonic mass distribution across all halos, but not how the different prescriptions alter the total baryonic masses (gas and star) within individual halos. We will address this question next.

\subsection{Total gas and stellar content in halos}
\label{sec:mass_content}

\begin{figure*}
	\includegraphics[width=\textwidth]{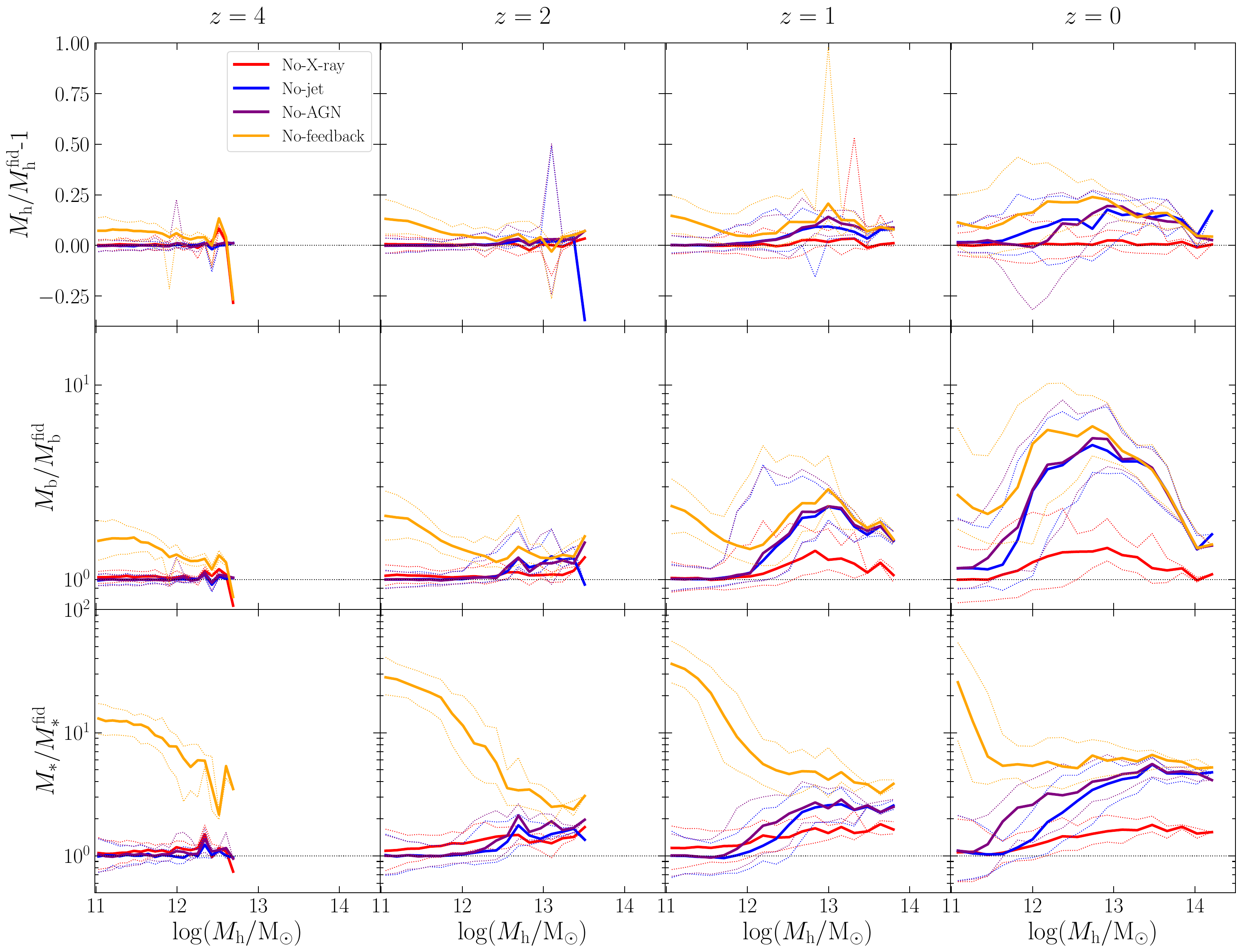}
    \caption{From the top to the bottom rows, relative difference of total, baryonic and stellar mass of halos at different redshift (as reported above every panel) and in different runs (colour coded as in Figure~\ref{fig:mfuncs_diff}), with respect to the fiducial-50 run. In the first row, the `relative difference' is shown on a linear scale, and is computed by taking the ratio of the total halo masses, and then subtracting one. In the other rows, the `relative difference' is defined simply as the mass ratio, and is plotted on a logarithmic scale. In all rows, the relative difference is computed by first selecting halos within a certain mass bin in the fiducial-50 run, and then seeking their counterparts in the other runs, defined as the halos that share the largest amount of DM particles (see main text for details). As such, the $x$-axis always refers to the total masses of halos in the fiducial-50 run. In all panels, solid lines refer to the median relative difference in each mass bin, while the dotted lines mark the 16$^{\rm th}$-84$^{\rm th}$ percentiles of the distribution. The total halo mass exhibits the smallest variations across all runs at all redshift. Conversely, the other components, and especially the stellar mass, can vary for more than one order of magnitude for the same halo across different runs. Therefore, when investigating the effect of feedback on halo properties, the halos should be selected by total halo mass rather than stellar mass.
    }
    \label{fig:mhalo_diff}
\end{figure*}

We now examine how feedback processes affect the mass content of individual halos. Because all $50 \hMpc$ boxes considered in this work start from the same initial conditions, we can identify the `copies' of the same halo across all runs. For this purpose, we can exploit the DM particle IDs, which are unique identifiers to each particle. We first read out the particle IDs of all DM particles associated to every halo in a given snapshot of the \simba\ $50 \hMpc$ run. We then do the same for the same snapshot of another \simba\ variant. At this point, we match every halo in the \simba\ $50 \hMpc$ simulation with the halo in the alternative run that shares the largest number of DM particle IDs with the original halo. In this sense, the halo in the target run represents a `copy' of  the halo in the original \simba\ $50 \hMpc$ simulation. 

We note that not every halo in a given run is necessarily matched to a copy in another \simba\ variant. As an example, let us consider the No-feedback run. We have already discussed in \S~\ref{sec:mass_funcs} that the absence of feedback processes favours halo growth. Indeed, the No-feedback run exhibits larger HMF and BMF even at the low-mass  end (see Figure~\ref{fig:mfuncs_diff}). Thus, there will be several halos in the No-feedback run that do not represent the copy of any halo in the \simba\ $50 \hMpc$. There are also numerical reasons why not all halos are paired with a counterpart in other runs. The FOF halo-finding algorithm that we adopted requires the linking of at least 32 particles for the creation of a halo object. Thus, even if only a few particles are moved beyond the linking length as a result of alternative feedback prescriptions, the smallest halos in a certain run may not find a counterpart in the other variants. In an even trickier scenario, two smaller halos may be associated to the same halo of a different run, if that halo ends up sharing enough particles with both of the original small halos. Therefore, the outcome of the halo-matching code may not always be invariant under permutations of the origin and target runs, especially for low-mass halos.

In order to avoid spurious matches, we take the a number of precautions. Whereas previously we considered halos containing less than 100 DM particles as poorly resolved, in this context we opt for a more conservative threshold of 1000 DM particles ($M_{\rm h}\approx 10^{11} \Msun$). Furthermore, we impose a minimum threshold on the percentage of shared particle IDs that two halos must have in order to be identified as a pair of  `halo copies'. We set such threshold to 90\%. This is very conservative, as we verified that even if we set it as low as 20\% there is still no significant increase of matches for low-mass halos. We verified that these two criteria produce identical halo pairs if we swap the origin and target runs, and as such the halo pairs can be considered genuine matches. 

After we run our halo-matching code as explained above, we organise the halos in the \simba\ $50 \hMpc$ into 20 logarithmic mass bins of equal width, within the halo mass range delimited by $10^{11} \Msun$ and the total halo mass of the largest halo in the snapshot considered. For every mass bin, we then compute the median, $16^{\rm th}$ and $84^{\rm th}$ percentile of the total mass distribution of the halo copies in the other runs. Finally, we calculate the relative difference of these statistics with respect to the central value of the halo mass bins defined earlier. This quantifies the statistical variation of the total mass of halos in the \simba\ $50 \hMpc$ due to different feedback prescriptions.

We show the variation in halo mass owing to baryonic physics in the top row of panels in Figure~\ref{fig:mhalo_diff}. Every panel refers to the redshift written in the upper part of the figure. The results of different runs are colour coded as in Figure~\ref{fig:mfuncs_diff}. The solid lines represent the median values, while the dotted lines indicate the $16^{\rm th} - 84^{\rm th}$ percentiles of the mass distribution, as explained earlier. The horizontal dotted line marks the zero level, to guide the eye. Overall, there is no significant variation on the halo mass $M_{200}$, at any redshift. The median relative difference between all runs and the \simba\ $50 \hMpc$ simulation is generally within $20\%$. The only exception is represented by the No-feedback run at $z=0$, where the relative difference can be as large as 40\%. As expected, variants with fewer active feedback modes exhibit larger differences with respect to the full-physics run, especially at $z=0$. In particular, we note that the masses of halos in the No-X-ray run match almost perfectly those of their copies in the \simba\ $50 \hMpc$ run. Furthermore, the No-AGN and No-jet runs display a similar mass distribution, suggesting that AGN jets play a more important role in altering the halo mass at low redshift than the AGN winds feedback mode does.

Following the same binning and matching procedure described above, in the mid-row of panels in Figure~\ref{fig:mhalo_diff} we plot the ratio of the baryonic mass of halos in the different \simba\ runs, with respect to the baryonic mass of their copies in the \simba\ $50 \hMpc$ simulation. The horizontal axis still shows the \textit{total} halo mass. The No-feedback run impacts the baryon mass within halos by up to a factor of $\sim 2$ down to $z=2$. The spread in the baryonic mass distribution is larger at $z=1$, and at $z=0$ the No-feedback run can introduce variations up to one order of magnitude in baryonic mass. Over the entire mass range considered, the halos in the No-feedback run have a larger mass than their counterparts in the \simba\ $50 \hMpc$. Instead, in the No-AGN and  No-jet runs, the baryon mass ratio with respect to the fiducial-50 simulation is consistent with one at $z \geq 1$ and $M_{\rm h} \lesssim 10^{12} \Msun$. In these variants, the largest differences occur at the intermediate halo mass at $z\leq 1$. At $z=0$, the baryonic mass of the halos in these runs can be a factor of $\sim 5-6$ larger than in the full-physics simulation. In contrast, the effect of X-ray feedback is much more confined to within halos, the largest relative differences with the \simba\ $50 \hMpc$ run being within $50\%$.

Finally, in the lower panels of Figure~\ref{fig:mhalo_diff} we show the ratio of the stellar mass of halos in different runs with respect to their copies in the \simba\ $50 \hMpc$ run, again as a function of the total halo mass. In this case, the differences between the No-feedback run and the \simba\ $50 \hMpc$ simulation are even more pronounced than for the baryonic mass. At $z=0$, such differences  reach a factor of $\sim7-8$ for $M_{\rm h} \gtrsim 10^{11.5} \Msun$, and rise up to a factor of $\sim 30$ for lower halo masses. This is a consequence of the overproduction of stars in the absence of any feedback mechanism. The other runs produce a visible excess of stellar mass for $z\leq2$. This is particularly evident at $z=0$, where the median ratio of the halo stellar masses in the No-AGN run with respect to the  \simba\ $50 \hMpc$ is $\sim 5$ at the high-mass end. The No-jet run yields analogous results, but the excess stellar mass appears only for $M_{\rm h} > 10^{12} \Msun$, whereas in the No-AGN run the overproduction of stars occurs already at $M_{\rm h} > 10^{11.3} \Msun$. This indicates that SN feedback dominates  galaxy quenching at lower halo mass, in agreement with the previous findings. The No-X-ray run exhibits smaller differences, always within a factor of two. For low-mass haloes ($M_{\rm h} \lesssim 10^{12} \Msun$), the results of the No-X-ray run are consistent with those of the fiducial simulation. This is presumably due to the bigger impact of X-rays on the BH kernel, which is generally larger for massive halos, and prevents the formation of stars in the innermost regions of the halos. We will verify this later when we consider the mass profiles within halos (see \S~\ref{sec:profiles}).

To sum up, Figure~\ref{fig:mfuncs_diff} tells us that whereas different feedback prescriptions greatly affect the baryonic and stellar mass of halos, especially at lower redshift, the halo mass $M_{200}$ is quite robust even under significantly different feedback prescriptions. This result has important implications for numerical works seeking to study the predictions of simulations with different feedback models on observables that are tied to halo properties. If no particle-based halo-matching technique is adopted, and halos are selected by mass, the most sound choice would be to utilise the total halo mass rather than the stellar mass. Operating a stellar-mass-based selection would risk comparing halos that may not constitute `halo copies' as explained in this section. As a consequence, it would be harder to understand which differences in halo properties other than the stellar mass are actually due to the different feedback prescriptions, or are somewhat spurious owing to the accidental comparison of completely separate halos across different runs. In particular, Figure~\ref{fig:mfuncs_diff} suggests that at redshift $z<2$ a selection by stellar mass can be seriously biased towards lower halo masses. 

Of course, we drew these conclusions based on the results of our suite of simulations, for which we do not re-calibrate feedback parameters such that they reproduce observations of key quantities such as the stellar mass function for all feedback variants (and indeed, such calibration is not obviously possible). In a suite of simulations where this is done, the variation of the halo stellar mass across different runs may be more limited. However, that would still need to be explicitly verified.

\subsection{Baryon abundances in different phases in halos}
\label{sec:omega}

In the previous section, we verified that the most sound choice for comparing halo properties across different feedback runs is selecting them by their total mass. Thus, we will now investigate the mass distribution of different baryonic phases as a function of the total halo mass and redshift, in all $50 \hMpc$ boxes. Our investigation represents an extension of \cite{Appleby_2021} work, who analysed the effect of feedback on the abundance of  different baryonic phases in $z=0$ \simba\ halos resembling the COS-Halos and COS-Dwarfs samples \citep{COS-Halos, COS-Dwarfs}. We will consider the same phases as \cite{Appleby_2021} did: hot, warm and cool CGM, wind, and ISM. 

\begin{figure*}
	\includegraphics[width=\textwidth]{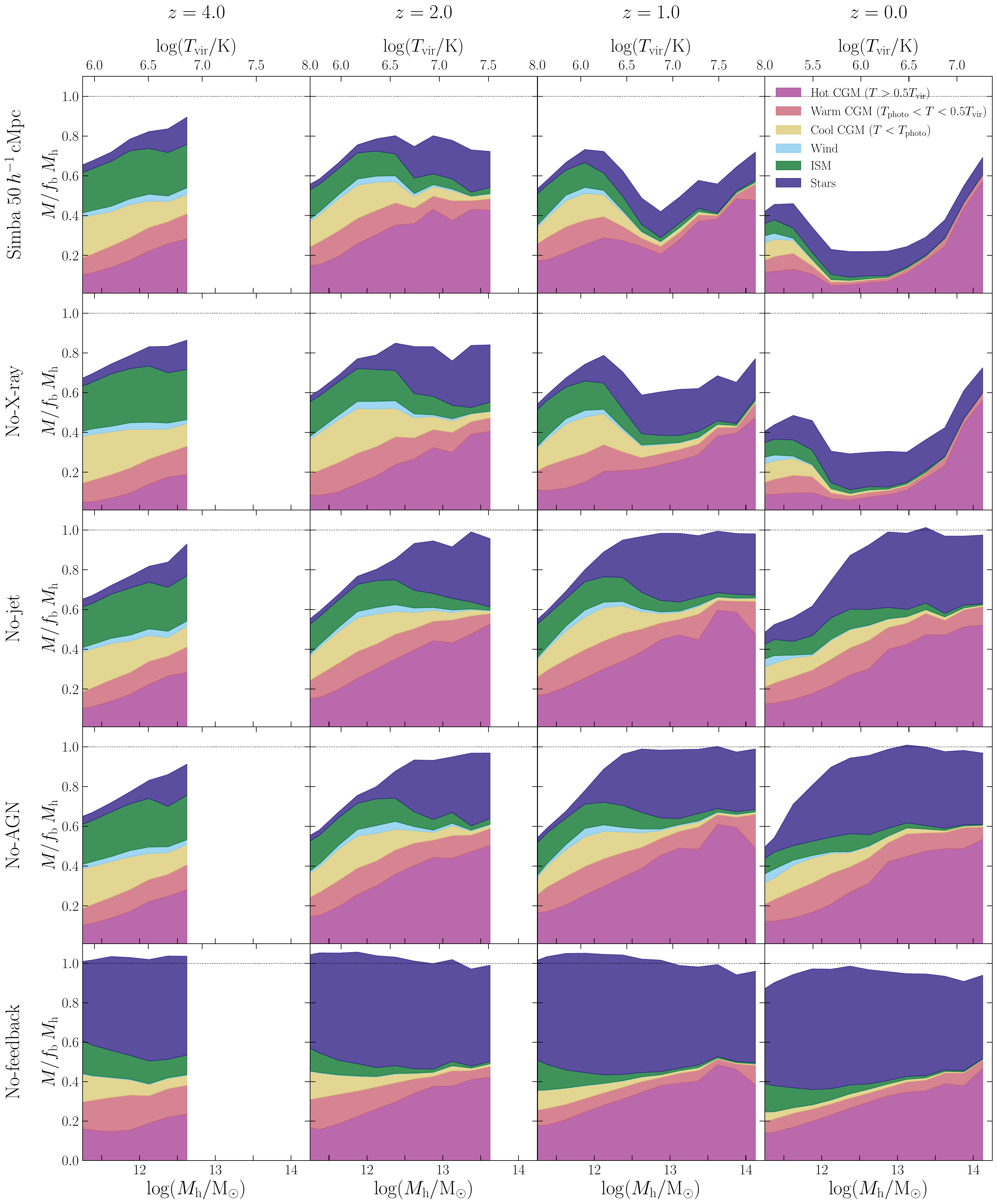}
    \caption{Median mass fraction of various baryonic components with respect to the cosmic share of baryon mass in halos, as a function of the total halo mass, for different \simba\ runs and different redshift. Each shaded area represents the contribution of different phases to the total baryon fraction, as indicated in the legend of the upper-right panel. Each row of panels refers to a different run, as reported on the left. Each column refers to a different redshift, as indicated at the top. In every panel, the upper $x$-axis reports the virial temperature (given by equation~\ref{eq:Tvir}) that corresponds to the halo masses (defined as $M_{200}$) indicated in the lower $x$-axis, at the redshift of interest. While in the No-feedback run the cosmic baryon fraction in halos is comparable to the cosmic baryon fraction $f_{\rm b}$, stellar feedback suppresses the baryon content of low-mass halos, and AGN jets are crucial to evacuate baryons from more massive halos at $z<2$.}
    \label{fig:mfrac_all}
\end{figure*}

We will also adopt the same definitions as in \cite{Appleby_2021}, here summarised in Table~\ref{tab:gas_phases}. We define ISM and wind particles following the criteria explained in \S~\ref{sec:simulations}. All other gas particles are split between the hot, warm and cool CGM phases. Specifically, gas particles with $T>0.5 \, T_{\rm vir}$, where $T_{\rm vir}$ is the virial temperature of the halo to which they belong, are considered `hot CGM' gas. The `cool CGM' is defined as gas with temperature below the photoionisation threshold ($T_{\rm photo} =10^{4.5} \, \rm K$), while the `warm CGM' phase is composed by gas particles with $T_{\rm photo} < T< 0.5 \, T_{\rm vir}$. The virial temperature is defined as in \cite{Mo_2002}:
\begin{equation}
\label{eq:Tvir}
    T_{\rm vir} = 3.6 \times 10^5 \, \left( \frac{V_{\rm c}}{100 \, \rm km \, s^{-1}} \right)^2 \, \rm  K \, ,
\end{equation}
where $V_{\rm c}$ is the circular velocity of the halo. The circular velocity was computed as $(G M_{200}/ r_{200})^{1/2}$, with $M_{200}$ and $r_{200}$ being the mass and radius corresponding to an enclosed mass density equal to 200 times the critical density. For a halo mass of $10^{11} \Msun$, we have $T_{\rm vir}\approx1.6 \times 10^5 \K$ at $z=0$.

\begin{table}
    \centering
     \caption{Definition of different gaseous phases considered in this work. See \S~\ref{sec:simulations} and \S~\ref{sec:omega} for details.}
    \begin{tabular}{cc}
    \hline
         Hot CGM &  $T>0.5 T_{\rm vir}$ \\
         Warm CGM & $T_{\rm photo}<T<0.5 T_{\rm vir}$\\
         Cool CGM & $T<T_{\rm photo}$\\
         Wind & hydrodynamically decoupled gas particles\\
         ISM & $n_{\rm H}>0.13, \rm cm^{-3}$ and\\  & $\log_{10}(T/\mathrm{K})<4.5+\log_{10}\left(n_{\rm H}/0.13 \, \rm cm^{-3}\right)$\\
         \hline
    \end{tabular}
    \label{tab:gas_phases}
\end{table}

We consider all well-resolved halos ($M_{\rm h} > 10^{11} \Msun$) in all $50 \hMpc$ boxes, and divide them into equally spaced logarithmic mass bins spanning the total halo mass range $10^{11} \Msun < M_{\rm h} < 10^{14.5} \Msun$ in increments of $0.25 \, \rm dex$. We note that in the halo mass range considered the condition $0.5 \, T_{\rm vir} > T_{\rm photo}$ is always satisfied, therefore the classification of the CGM phases presented in Table~\ref{tab:gas_phases} is always well defined. We then compute the median mass of the aforementioned gaseous phases and of the median stellar mass contained in all halos within each bin, normalised by $f_{\rm b} M_{\rm h}$. We plot the results of our analysis in Figure~\ref{fig:mfrac_all}. All panels in the same row refer to the same \simba\ run, as specified in the left part of the figure. The panels in the same column refer to the same snapshot, corresponding to the redshift written in the upper part of the figure. The upper $x$-axis reports the virial temperature that corresponds to the halo mass at the redshift in question, following equation~\eqref{eq:Tvir}. In all panels, we plot the median mass fraction of all phases with shaded areas, which are colour coded as specified in the legend inside the upper-right panel. By definition, when the cumulative mass fraction of all phases in a certain halo mass bin is equal to one, then the total baryon mass fraction inside the halos in the bin in question is equal to the cosmic baryon mass fraction. For this reason, we include a horizontal dotted line that corresponds to the cosmic baryon mass fraction in all panels.

The No-feedback variant exhibits strikingly different results compared with any other run. Almost all halos have their cosmic share of baryonic mass, and the mass fraction of stars is almost constant throughout all redshifts and mass bins. This is a direct consequence of the absence of any feedback prescriptions, which promotes star formation and does not prevent baryons from accreting onto halos. The warm and hot CGM phases dominate over the other gaseous phases at the high-mass end for $z<2$: since there is no additional source of heating, this is likely caused by shock heating or photoionisation of gas with $T>T_{\rm photo}$ induced by the UV background. At later times, halos have lower characteristic densities, and thus cooling times become larger. Therefore, gas cooling is less efficient, and this would explain the larger share of warm and hot CGM gas over cool CGM and ISM in large enough haloes ($M_{\rm h} \gtrsim 10^{12} \Msun$). In lower mass haloes, the relative abundance of cooler and warmer phases is approximately equal even at $z=0$.

As expected, the activation of stellar feedback suppresses star formation. Thus, with respect to the No-feedback run, we find a lower stellar mass fraction for lower-mass halos. Conversely, the mass fraction of the gaseous phases increases. Another consequence of stellar feedback is the decrease of the total baryon fraction in low-mass halos. At $z=4$, the baryon mass fraction is correlated to the total halo mass in the form of a power law. At $z \leq 2$, we can clearly see that the power law saturates at high enough masses. This behaviour is reminiscent of the baryonic Tully-Fisher relationship (bTFR), i.e. the empirical power-law correlation between baryonic and total halo mass, below a certain critical halo mass; above the critical mass, haloes retain their cosmic share of baryons \citep[see, e.g.,][]{McGaugh_2010}. In the No-AGN run we find that, at $z=0$, such critical mass is $M\approx 6\times 10^{12} \Msun$. This value is of the same order of the knee of the bTFR found by \cite{McGaugh_2010} in a compilation of observations of a variety of galaxies and clusters (see their figure~1). Although it is interesting to see that stellar feedback alone can in principle give rise to a feature resembling the bTFR (see also the appendix of \citealt{SP21}), a direct comparison between the bTFR in \simba\ and observations is beyond the scope of this work. In fact, this was the subject of \cite{Glowacki_2020}, who considered a set of \simba\ galaxies that reflects the characteristics of the samples in observations (see also \citealt{Glowacki_2021}). Instead, in Figure~\ref{fig:mfrac_all} we are considering all halos with $M>10^{11} \Msun$.

Our results for the No-AGN run are also broadly in agreement with an analogous study by \cite{Davies_2019}, who analysed the baryon mass fraction in galaxies within the EAGLE simulation \citep{EAGLE_Schaye2015}, and in a variant without AGN feedback. They find that in the No-AGN run the baryon mass fraction saturates to $f_{\rm b}$ for halos with $M_{200} \gtrsim 10^{12.5} \Msun$, and exhibits a power-law behaviour at lower masses. Hence, \simba\ exhibits qualitatively the same behaviour, although we find that the median baryon mass fraction at $M \approx 10^{11.5} \Msun$ is around 50\% at present time, whereas \cite{Davies_2019} obtain a value around 30\%. However,  \cite{Davies_2019} also find a large scatter around the median value, with some haloes containing a baryonic mass fraction as large as 60\%.

In Figure~\ref{fig:mfrac_all}, we can see that switching on AGN radiative winds only mildly affects the mass split between stellar and gaseous phases at all redshifts, and does not qualitatively impact the overall trends discussed for the No-AGN run. However, we do notice a larger amount of hot CGM gas at all redshift, and throughout the entire halo mass range considered, putatively owing to AGN wind energy deposition. Another difference with respect the No-AGN run is that the bTFR seems to be slightly steeper, and the knee of the bTFR occurs at higher masses.

AGN jets appear to be the real game changer. In the No-X-ray run, we notice a suppression of stars and all gaseous phases except the hot CGM gas at $z=1$ in the most massive halos ($M\gtrsim 10^{13.5} \, \rm \Msun$, i.e. clusters. At $z=0$, these are the only halos that retain more than 50\% of their cosmic share of baryons. Only halos with $M_{\rm h} < 10^{12} \Msun$ contain an appreciable mass of ISM, cool/warm CGM and winds. Otherwise, the content of halos is essentially split between stars and hot CGM. Conversely, at higher redshift ($z\geq2$), the results of the No-X-ray run are very similar to those of the No-AGN and No-jet runs. Once again, our results support the thesis that AGN jets are the dominant feedback mechanism impacting the baryon content of halos at $z < 2$, whereas at higher redshift stellar feedback is the primary physical process in this respect \citep[see also][]{Christiansen_2020, Sorini_2020}. This is a consequence of the fact that AGN jets become more ubiquitous in massive galaxies at $z\la 2$, when the central BHs of the most massive halos have grown enough to trigger this feedback mode. 

Our results for the No-jet run are qualitatively similar to those found by \cite{Davies_2019} in the fiducial run of the EAGLE simulation. However, the AGN feedback model in the EAGLE simulation is based on a single mechanism that transfers part of the energy of gas accreting on to BHs to the surrounding gas, hence increasing its temperature. This is thus different from the tri-modal AGN feedback prescription in \simba. Thus, a one-to-one comparison between the results of the two simulations as far as the baryon mass fraction in halos is concerned is not straightforward. Nevertheless, it is noteworthy that both EAGLE and \simba\ suggest that AGN feedback mechanisms are crucial to evacuate baryons from halos at low redshift. In a follow-up work, \cite{Oppenheimer_2020} showed that the mass fraction of baryons in the CGM of $L_*$ galaxies in the EAGLE simulation is anti-correlated to the mass of the central BH, arguing that more massive BHs can transfer more energy to the surrounding baryons, and drive them outside the virial radius \citep[see also][]{Davies_2020, Davies_2021}. This scenario was later substantiated with zoom-in simulations too \citep{Davies_2022}, and lends support to our interpretation of the results of the \simba\ No-X-ray run just discussed.

Finally, we notice that adding X-ray heating does not qualitatively change the results of the No-X-ray run. The most notable difference is the slightly lower stellar mass fraction for $M_{\rm h} >10^{12} \Msun$ at $z\leq1$. This may be explained by the fact that X-ray heating additionally quenches star formation around the BH kernel. If a large stellar mass is concentrated in the central regions of the most massive halos, then X-ray heating is expected to make a visible difference. Clearly, we cannot verify this hypothesis from the analysis of Figure~\ref{fig:mfrac_all}, which is agnostic with respect to the spatial mass distribution within single halos. To obtain this information, we should look into the density and mass profiles within individual halos. This will be the main topic of the next section.

\section{Mass radial distribution within halos}
\label{sec:profiles}

\begin{figure*}
    \centering
    \includegraphics[width=\textwidth]{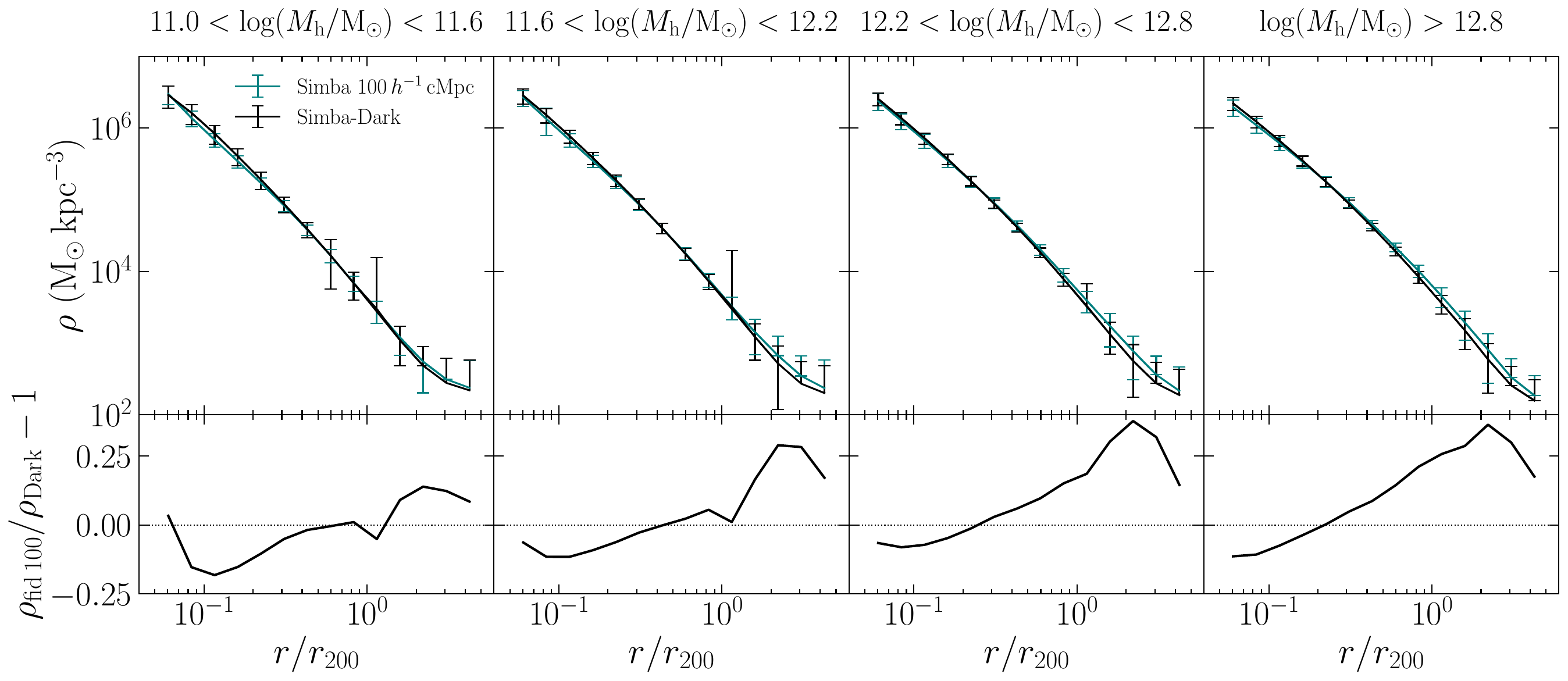}
    \caption{\textit{Upper panels}: Dark matter mass density profiles of halos within the total mass bin indicated at the top, as a function of the radial distance from the centre of the halos, normalised by the virial radius $r_{200}$. The solid lines represent the median mass density in each radial distance bin, and the error bars the standard deviation of the distribution within such bins. The teal and black density profiles refer to the $z=0$ snapshot of the \simba\ $100 \hMpc$ and \simba-Dark runs, respectively. \textit{Lower panels}: Relative difference between the halo dark matter profiles in the \simba-Dark and the \simba\ $100 \hMpc$ runs. The inclusion of baryonic physics is seen to lower the dark matter density towards the centres of halos, while increasing its density in the vicinity of halos, more so towards higher halo masses.}
    \label{fig:dprof_dark}
\end{figure*}

In this section, we will study the effect of baryons on the internal structure of halos. To begin with, we will investigate how the presence of baryons affects the dark matter density profile in \simba\, by comparing the fiducial-100 run with the DMO variant. We will then analyse the density profiles of different baryonic phases in all \simba\ runs with different feedback prescriptions. We will then determine the distance at which the baryon mass fraction of a halo equals the cosmic baryon mass fraction.

\subsection{Density profiles}
\label{sec:dprof}

It is well known that the DM density profiles of halos in N-body simulations approximately follow a universal scale-invariant mass density profile, which is well described by the Navarro-Frenk-White (NFW) profile \citep{NFW}. Such profile is characterised by a cusp in the innermost regions of the halo, but as discussed in the introduction, hydrodynamic simulations manage to smooth out this feature. At larger radii, hydrodynamic simulations tend to agree with their DMO counterparts \citep{Mashchenko_2008, Madau_2014, Oman_2015}.

Here we quantify this behaviour in the \simba\ simulation. We select all halos in the $z=0$ snapshot of the fiducial-100 run, and consider all DM particles between $0.01 r_{200}$ and $5r_{200}$ from the point of minimum gravitational potential in each halo. We then divide them into 20 logarithmic bins of distance of equal width. In this way, we can immediately obtain the total DM mass within each radial shell, and straightforwardly compute the density profile. We further separate the halos into four groups based on their total mass: $10^{11} - 10^{11.6} \Msun$, $10^{11.6} - 10^{12.2} \Msun$, $10^{12.2} - 10^{12.8} \Msun$, and $M_{\rm h}>10^{12.8} \Msun$. We can then take the density profiles of the halos within each group and compute the mean density for every distance bin (in units of $r_{200}$). 

The results are plotted with the teal lines in Figure~\ref{fig:dprof_dark}. Every panel reports the results for the total halo mass bin written in the upper part of the figure. The error bars represent the standard deviation of the DM density distribution within each distance bin. For ease of representation, we decided to plot the full error bar only when its lower bound is above the lower limit of the $y$-axis. In the rare occasions when this is not the case, we then plot only the upper error bar. We also computed the scatter of the DM density profile due to cosmic variance, following the same procedure described in \S~\ref{sec:mass_funcs} for the mass functions. This is negligible with respect to the to halo-to-halo scatter, therefore we do not plot it.

We repeat the same procedure described above for the \simba-Dark run too. The results are plotted in Figure~\ref{fig:dprof_dark} with black lines. The DM density is divided by a corrective factor of $(1+f_{\rm b})$, as the mass of DM particles in the \simba-Dark simulation is larger than in the fiducial-100 run, to compensate for the mass of the baryons that are not included in the simulation. Every small panel reports the relative difference between the DM density profile in the fiducial-100 simulation with respect to \simba-Dark in the halo mass bin shown in the large panel above. The horizontal dotted line corresponds to a null difference, to guide the eye. We restrict the $x$-axis to the range $0.05-5\, r_{200}$, as we verified that the density profiles are not well converged resolution-wise for $r<0.05\,r_{200}$ (see Appendix~\ref{sec:convergence}). Overall, the fiducial-100 and \simba-Dark runs predict similar DM density profiles, with relative differences smaller than $25\%$ within the virial radius. Interestingly, the addition of baryons increases the DM density profile outside the virial radius, increasing it by almost $\sim 40\%$ at $r \sim 2\, r_{200}$ in higher-mass halos ($M_{\rm h} >10^{12.2} \Msun$).  We found qualitatively similar results also at higher redshift ($z=1$, $z=2$ and $z=4$). Unfortunately, the poor convergence at $r<0.05\,r_{200}$ does not allow us to conclusively determine whether halos in the fiducial-100 run exhibit less cuspy profiles than in its DMO counterpart. To undertake this study, we would need a higher-resolution simulation, and we leave this for future work.

Having analysed the overall effect of baryons on the DM profiles, we can now look into the density profile of the baryonic components of halos in the \simba\ $100 \hMpc$ run. We once again separate halos into bins of total mass, and organise gas and stellar particles within radial shells, as we have described earlier for the DM profiles. We can then take our analysis further, by classifying the gas particles among the different phases defined in \S~\ref{sec:omega}, hence obtaining the density profile for each of such phases.

\begin{figure*}
    \centering
    \includegraphics[width=\textwidth]{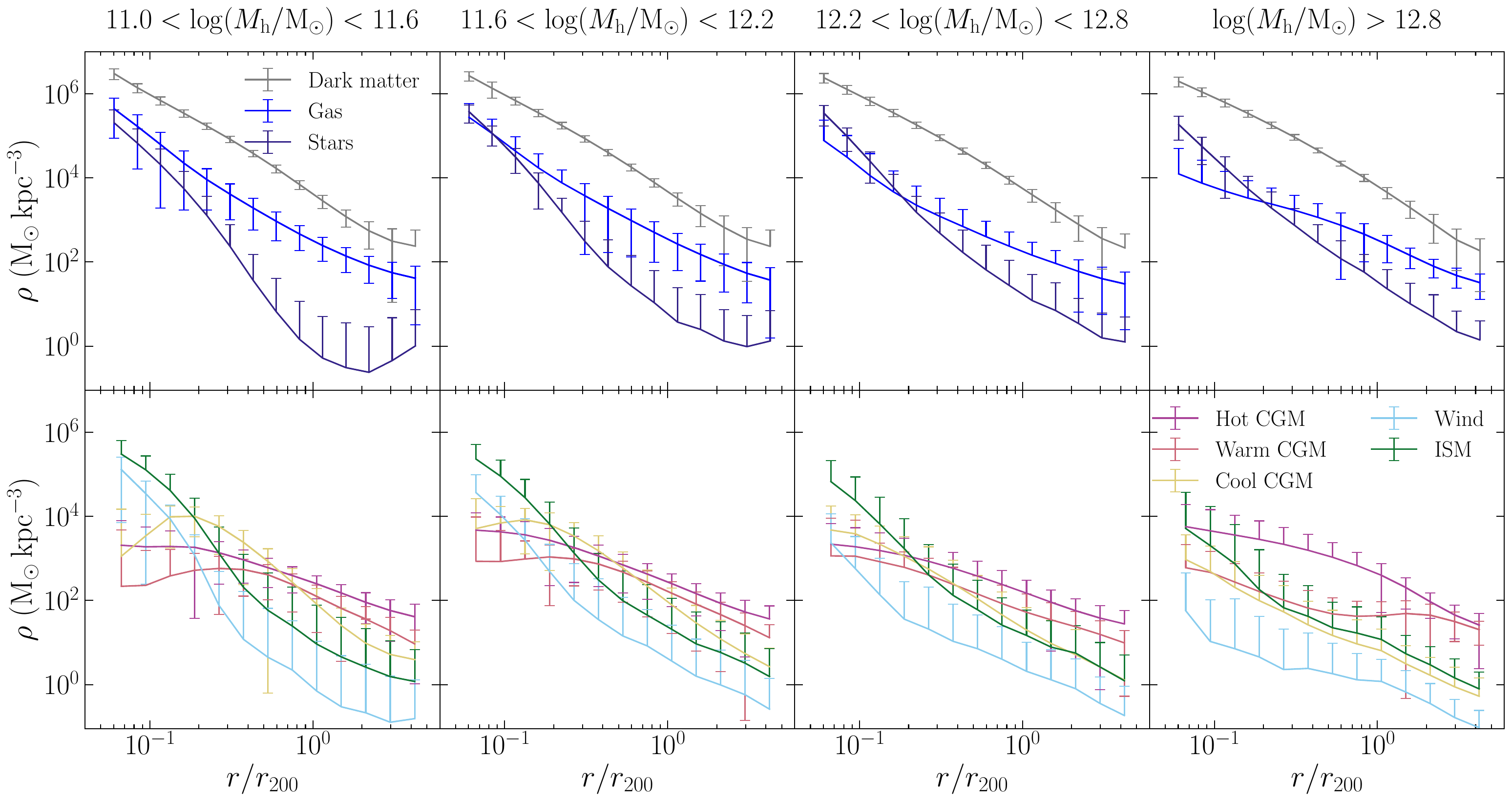}
    \caption{\textit{Upper panels}: Dark matter, gas and stellar mass density profiles (grey, blue and purple lines, respectively) of halos within the total mass bin indicated at the top, as a function of the radial distance from the centre of the halos, normalised by the virial radius $r_{200}$. The solid lines represent the mean mass density in each radial distance bin, and the error bars the standard deviation distribution within such bins. All density profiles refer to the $z=0$ snapshot of the \simba\ $100 \hMpc$ run. \textit{Lower panels}: Same as in the upper panels, but for different gaseous phases, following the same colour coding as in Figure~\ref{fig:mfrac_all}. Gas overall traces the dark matter profile, but stars and ISM gas in particular are more centrally concentrated, while hotter CGM components dominate in the outskirts increasingly so at higher halo masses.}
    \label{fig:dprof}
\end{figure*}

In the upper panels of Figure~\ref{fig:dprof} we plot the density profile of stars and all gas elements within halos in the $z=0$ snapshot of the \simba\ $100 \hMpc$ simulation (blue and purple lines, respectively). As a reference, we plot again the DM density profile with a grey line. The error bars represent the standard deviation of the density profiles of the halos within each mass bin, and are drawn using the same conventions that we adopted for Figure~\ref{fig:dprof_dark}.

As a general trend, we notice that the stellar component progressively dominates over the gaseous components in the core of the halos as the halo mass increases, while the overall gas density decreases. In the lowest mass bin the gaseous component dominates over stars, especially in the outer regions of the halo. Around the virial radius, the gas density is more than $2 \, \rm dex$ larger than the stellar density, indicating the inefficient conversion of gas into stars owing to strong stellar feedback in such systems. Beyond $2 \, r_{200}$, the stellar mass density profile exhibits an upturn, which is probably caused by the presence of nearby halos. In the halo mass bin $10^{11.6}<M_{\rm h}/\Msun <10^{12.2}$, the stellar density profile follows the one of the gaseous component up to $\sim 0.1 r_{200}$; beyond this radius, stars are sub-dominant with respect to gas. For halos with $M_{\rm h} > 10^{12.2} \Msun$, the stellar density profile is peaked to higher values in the central region, and decreases steeply after $0.2 \, r_{200}$, beyond which gas dominates over stars. This is consistent with the expectation from AGN feedback activity, which would move gas towards the outskirts of massive halos and quench star formation. In the largest mass bin, the gas density profile in the inner portions has actually been even reduced compared to the lowest halo mass bin, indicative of gas evacuation. We can further subdivide the gaseous component into various phases. In the lower panels of Figure~\ref{fig:dprof} we show the density profiles of the different gaseous phases, as defined in  \S~\ref{sec:omega}. We use the same colour coding as in Figure~\ref{fig:mfrac_all}. 

For all halos, ISM gas dominates the inner regions but drops off more quickly than the CGM components, mimicking the stellar profile.  The wind component tracks the ISM gas, and drops in amplitude more quickly than ISM gas in higher mass halos, reflecting the lower mass loading factors at high masses.  Also towards higher masses, the warmer CGM gas phases dominate over the cooler ones. In particular, in the highest mass bin, hot CGM gas constitutes the dominant contribution to the total gas density profile for $r>0.1 \, r_{200}$ and out to $r=5 r_{200}$. This behaviour owes to a combination of virial shock heating \citep{Dekel_2003,Keres_2005,Gabor_2012} along with AGN jet feedback that pushes CGM gas outside of halos.

To validate our hypotheses on the role of different feedback prescriptions on the various baryon components in the density profiles shown in Figure~\ref{fig:dprof}, we need to repeat our analysis for every feedback variant of \simba. Thus, the baseline model will be \simba\ $50 \hMpc$ from now on. We have verified that all density profiles are well converged volume-wise in the entire radial distance range and in all mass bins considered (see the Appendix~\ref{sec:convergence}). Therefore, the density profiles of the \simba\ $50 \hMpc$ run are statistically indistinguishable from those of the \simba\ $100 \hMpc$ run, already shown in Figure~\ref{fig:dprof}.

\begin{figure*}
    \centering
    \includegraphics[width=\textwidth]{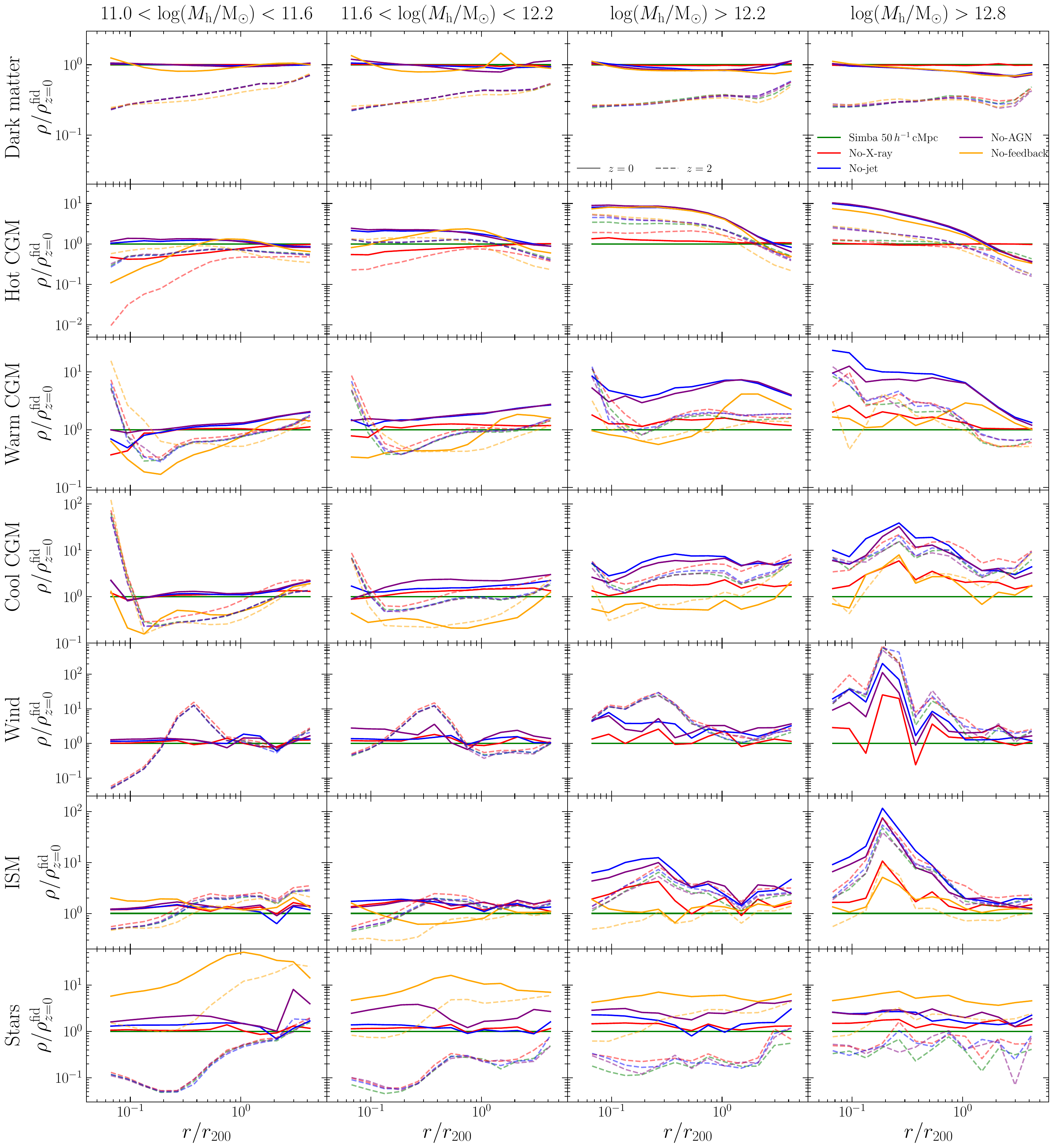}
    \caption{Ratio between the mean comoving mass density profile (obtained as explained in \S~\ref{sec:dprof}) of different components within halos, with respect to the results of the \simba\ $50 \hMpc$ run at $z=0$. In all panels, the $x$-axis shows the radial distance from the centre of the halos, normalised by $r_{200}$. Each column refers to halos within the total mass bin indicated at the top. Each row shows the results for a different component, as reported at the left. In every panel, the green, red, blue, purple, and orange lines represent the results from the \simba\ $50 \hMpc$, No-X-ray, No-jet, No-AGN and No-feedback runs, respectively. The solid,and dashed lines refer to the snapshots at $z=0$ and $z=2$, respectively.}
    \label{fig:dprof_diff}
\end{figure*}

Figure~\ref{fig:dprof_diff} summarises our analysis on the effect of feedback prescriptions on the density profiles of different components of halos, as a function of their mass and redshift. Every row reports the results for a different component, as specified in the left part of the figure. Every column refers to the density profiles of the halos within the mass bin specified in the upper part of the figure. In all panels, we show the ratio of the \textit{comoving} density profile in every run with respect to the density profile given by the \simba\ $50 \hMpc$ run at $z=0$. The colour of each line corresponds to a different feedback variant, as specified in the legend inside the upper-left panel. Solid and dashed lines refer to results at $z=0$ and $z=2$, respectively. Because we are plotting the ratios of comoving density profiles, the redshift-evolution is due to astrophysics only, and does not incorporate the effect of the expansion of the Universe.

Clearly, Figure~\ref{fig:dprof_diff} encodes a considerable amount of information. In discussing it, we will focus on the results that are in our view most noteworthy. First of all, we notice that, for a fixed redshift, the DM profiles in all runs agree within $\sim 20\%$, which are thus profiles are very robust to changes in baryonic physics. This is consistent with what we already found for the HMF and the total mass content of individual halos in \S~\ref{sec:mass_funcs}-\ref{sec:mass_content}.

In runs without AGN jet feedback, high-mass halos ($M_{\rm h} > 10^{12.2} \, \rm M_{\odot}$) contain almost ten times as much hot CGM gas within $0.1 r_{200}$ with respect to the fiducial-50 run. As jets are activated, the hot CGM gas density profile within the virial radius is lower at $z=0$. The observed trend unequivocally confirms that AGN jets are the main driver of the lowered hot gas density distribution inside and around high-mass halos at low redshift. Conversely, at $z=0$ in low-mass halos ($M_{\rm h} < 10^{12.2}\, \rm M_{\odot}$) most runs exhibit much smaller relative differences, within a factor of $3$ at $r>0.1\, r_{200}$. This means that in low-mass halos the effect of AGN jets is much less important. Indeed, the halos in question are unlikely to host massive BHs which can trigger AGN jets, hence they are not able to expel hot gas as effectively as their high-mass counterparts. On the other hand, it would seem that X-ray heating has a stronger effect: switching this feedback mode off would reduce by a factor of $\sim2-3$ the amount of hot CGM at low redshift, and even by one order of magnitude at $z=2$ for $r \lesssim 0.2 \, r_{200}$. This may seem a somewhat surprising result, as only a few halos are eligible for X-ray feedback at $z \gtrsim 2$ in \simba\ $50 \hMpc$ as per the criteria explained in \S~\ref{sec:simulations} \citep{Sorini_2020}. However, hot CGM gas accounts for $<2\%$ of the total baryonic mass enclosed within $0.2 \, r_{200}$ of halos in the lowest mass bin (Figure~\ref{fig:Omprof}). Thus, even the X-ray heating generated by a few halos may be sufficient to produce a large \textit{relative} difference in the hot CGM density profile within the halo core. Therefore, this feature may not be statistically significant, and a larger simulation would be required to explicitly test that.

The impact of feedback on the density profile of the warm CGM is largest within $r<0.1\, r_{200}$, i.e. where the density profiles are less converged (see Appendix~\ref{sec:convergence}). Beyond this radius, the density of warm CGM increases with redshift for $M_{\rm h}< 10^{12.2} \Msun$, while it decreases at late times for $M_{\rm h}>10^{12.2} \Msun$. Furthermore, for $r>0.1\, r_{200}$, runs incorporating more feedback modes generate lower warm CGM densities at low redshift, for all mass bins. Such trend is more evident in higher masses. We interpret the observed trends as a result of the different intensity of the various feedback mechanisms in halos of different mass. In low-mass halos, supernovae-driven winds tend to heat gas, hence building up warm CGM over time. On the other hand, in high-mass halos AGN jets are more effective at sweeping the excess warm gas outside halos. We also notice that X-ray feedback has a strong effect on the warm CGM profile within $0.1\, r_{200}$ in the highest-mass bin, reducing the warm CGM density by about a factor of $\sim 2$. This is likely a consequence of the kinetic component of X-ray feedback, which adds extra energy to the gas surrounding the central BH kernel, hence allowing it to move towards larger radii.

The amount of cool CGM is relatively unaffected by AGN feedback $z\geq 2$, in all halo mass bins. The ratios between the density profiles of two different \simba\ variants stays always within a factor of $\sim 2$. Suppressing also stellar feedback can introduce differences of up to an order of magnitude in the density profile. At $z=0$, the addition of feedback mechanisms suppresses the density of cool CGM gas everywhere in the halo. The No-feedback run exhibits a somewhat different behaviour, though. While at $z=2$ it overproduces cool gas within the halo core, it produces less cool gas than in the \simba\ $50 \hMpc$ run in the region $r \lesssim 0.1\, r_{200}$, for halos with $M_{\rm h}<10^{12.2} \Msun$. In other words, at high redshift cool gas is more concentrated in the innermost regions of the halo in the No-feedback run. At $z=0$, the cool gas density profile is actually smaller than in the fiducial-50 run. On the other hand, in the highest mass bin the cool CGM gas is more evenly distributed, both at $z=2$ and $z=0$, and its density profile appears to be systematically larger than in the \simba\ $50 \hMpc$ simulation at $z=0$. Also in this case, X-ray heating causes a significant suppression on the density profile of cool CGM. The reason is probably the volume heating component of X-ray feedback, which increases the temperature of the gas within the kernel of the central BH, hence diminishing the supply of cool gas.

For halos with $M_{\rm h} <10^{12.2} \Msun$, the density profile of gas particles in the wind phase is only minimally affected by feedback. In the mass bin $10^{12.2}-10^{12.8} \Msun$, more significant differences appear, albeit only at $z=0$. In the highest-mass bin, the effect of AGN jets and X-ray heating have a strong impact on the wind density profile at $z=0$. The observed trends reflect the fact that AGN feedback modes are barely active at $z>2$, and inefficient for $M_{\rm h}< 10^{12.2}\Msun$. In more massive halos at $z=0$, AGN-driven feedback is more effective at removing gas particles from the halos, including wind particles.

In all runs, ISM gas is more scarce in the central regions ($r<0.1\, r_{200}$) of $M_h < 10^{12.2} \Msun$ halos at higher redshift. Over time, the production of stars activates a feedback mechanism that heats up gas, thus diminishing the amount of ISM in favour of hot CGM gas in the fiducial-50 run. This is consistent with what we found for the mass fraction of different gaseous phases as a function of the total halo mass (see \S~\ref{sec:omega}). However, at higher halo masses the runs with at least stellar feedback exhibit a  larger ISM density than in the fiducial run for $r<0.1 r_{200}$. At $z=0$ and $M_{\rm h} > 10^{12.2} \Msun$, the No-X-ray simulation produces less ISM gas with respect to the No-AGN and No-jet runs because of the enhanced jet-driven heating. The No-feedback run also yields less ISM gas, but in this case the reason is that the ISM is consumed too efficiently from new stars. 

The No-feedback run produces indeed larger star densities at all redshifts, mass bins, and radii, with respect to the fiducial-50 run. Stars are not confined into the inner regions of the halos only, but rather extend all over the halo. The suppression of star formation is evident in all other runs. At $z=0$, jets are again determinant in diminishing star formation. The No-X-ray run differs by a factor of $\sim 2$ from the \simba\ $50 \hMpc$ run. This is because the extra X-ray heating contributes to quench star formation, especially in the central regions of the most massive halos, as discussed earlier. In the intermediate mass range $10^{11.6} - 10^{12.8} \Msun$, AGN winds have a more significant impact on the density profile than for the other cases, introducing a relative difference of a factor of $2-3$ with respect to the No-AGN run. However, we caution that the convergence in the stellar and ISM density profiles with respect to mass resolution is not optimal (see Appendix~\ref{sec:convergence}).

One of the main take-home messages of this extended analysis is that AGN jets play a key role in shaping halo gas distribution in the \simba\ simulation for $z<2$. It would be interesting to understand up to what radius they can extend their influence, as a function of halo mass and redshift. This is the question that we will address next.

\subsection{Enclosed mass profiles}
\label{sec:mass_prof}

\begin{figure*}
    \centering
    \includegraphics[width=\textwidth]{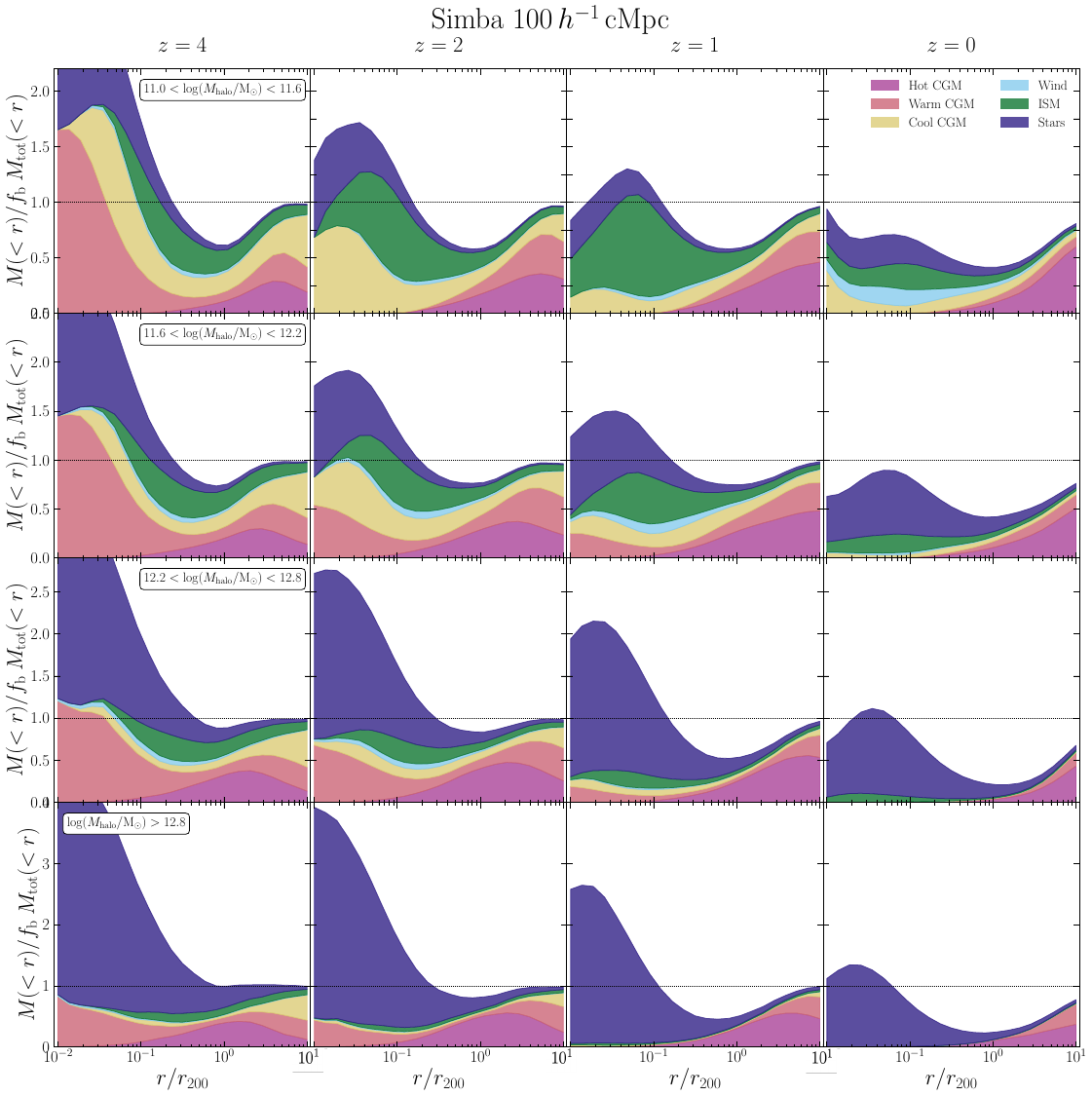}
    \caption{Median mass of different components enclosed within different distances from the centre of halos within different mass bins, normalised by the expected baryonic mass within the same distances under the assumption that the baryon mass fraction equals the cosmic value $f_{\rm b} = \Omega_{\rm b}/\Omega_{\rm m}$. Each row corresponds to a different total halo mass bin, as indicated inside the first panel on the left in each row. Each column refers to a different redshift, as reported at the top. In all panels, the $x$-axis shows the radial distance from the centre of the halos, normalised by $r_{200}$. The colour coding is the same as in Figure~\ref{fig:mfrac_all}. The horizontal dashed line in all panels marks the cosmic baryon mass fraction. At high redshift, rapid cooling leads to a concentration of gas and stars towards the centres, while at lower redshifts halos become increasingly evacuated of baryons in their outskirts and surroundings.}
    \label{fig:Omprof}
\end{figure*}

We now expand the analysis presented in the previous section by investigating the radial profile of the enclosed baryon mass within and around halos. We consider all halos in the \simba\ $100 \hMpc$ run, and then define a set of 28 spheres centred at the point of minimum gravitational potential. The radii of the spheres are chosen such that they span the range $0.02-50 \, r_{200}$, with equal logarithmic increments. We then compute the mass of star particles and gas elements enclosed within every sphere. We classify the gas particles into the same phases considered in \S~\ref{sec:omega} and \S~\ref{sec:dprof}. Splitting the halos into the same bins of total mass defined in \S~\ref{sec:dprof}, we can then easily obtain the enclosed mass profile of different phases around halos of different mass.

Figure~\ref{fig:Omprof} shows our results for the \simba\ $100 \hMpc$ run. The profiles within every panel are normalised to the total baryon mass that would be enclosed in each of the aforementioned spheres, if the baryon mass fraction within the corresponding radius where equal to the cosmic baryon mass fraction. The horizontal dotted line refers to an enclosed mass fraction equal to $f_{\rm b}$. The contributions of each phase to the total baryon mass fraction is represented with the same colour coding as in Figure~\ref{fig:mfrac_all}. All panels in the same row refer to the same halo mass bin, as indicated in the left-most panel. Along a given row, each panel refers to a different redshift, as indicated in the upper part of the figure. 

Overall, at higher redshift there is a higher fraction of  stars, ISM and cool CGM, which extend even up to tens of virial radii from the centres of halos. As time goes by, the warm/hot CGM phases gradually dominate, most significantly at $r> r_{200}$. This is a consequence of both feedback mechanisms and the lowered density, which makes radiative cooling less effective at counteracting shock heating. At $z=0$, warm/hot CGM gas dominates outside the virial radius. This is  consistent with the large amount of WHIM gas in the IGM, shown in  Figure~\ref{fig:fIGM}. For halos with mass $M_{\rm h} >10^{12.2}\Msun$, at $z=0$ the mass enclosed in the region within the virial radius is almost entirely made of stars, while in the lowest-mass bin there is a comparably large fraction of cool CGM, ISM and wind particles. For all mass bins, stars are more concentrated in the halo core at high redshift. At later times, stars are quenched in the central regions of halos, while their contribution to the total baryon mass fraction raises out to larger radii. 

We notice that the distance at which the total enclosed baryon mass fraction (besides the halo central regions) saturates to the cosmic baryon mass fraction is always beyond one virial radius, with the only exception of massive halos $(M_{\rm h}> 10^{12.8}\Msun$) at $z=4$. At $z< 4$, all halos are baryon deficient at $r \sim r_{200}$. Also, such distance increases as redshift decreases. At $z=0$, for the higher mass bins, it is even $>10 \, r_{200}$. To understand what causes this noteworthy feature, we will now investigate how such `critical radius' changes if we switch off one or more of the feedback prescriptions. For this purpose, we need to compare all feedback variants with the \simba\ $50 \hMpc$ run, which stars from the same initial conditions as all other $50 \hMpc$ boxes. We verified that the median enclosed mass profiles in the \simba\ $50 \hMpc$ run are almost indistinguishable from those of the \simba\ $100 \hMpc$, shown in Figure~\ref{fig:Omprof}.

\begin{figure}
    \centering
    \includegraphics[width=\columnwidth]{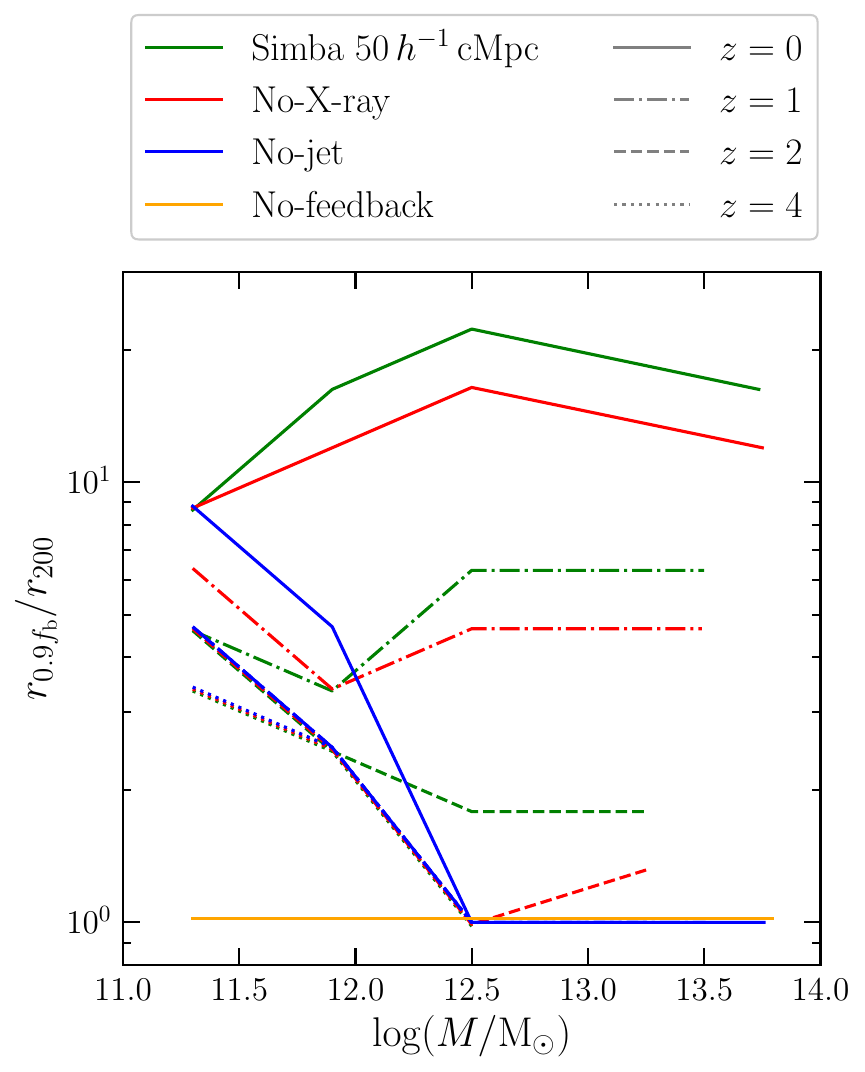}
    \caption{Distance $r_{0.9\, f_{\rm b}}$ from the centre of the halo where the enclosed baryonic mass density equals 90\% of the average baryonic mass density, as a function of the total halo mass. Each line represents the median $r_{0.9\, f_{\rm b}}$ within the same total halo mass bins considered in Figure~\ref{fig:Omprof}. The colour coding and line styles are the same as in Figure~\ref{fig:dprof_diff}. To aid the readability of the plot, we omitted the No-AGN run, as it gives the same results as the No-jet run. Jets are crucial to evacuate baryons from halos with mass $M>10^{12}\, \rm M_{\odot}$ at $z<2$, while stellar feedback is responsible for evacuating low-mass halos. }
    \label{fig:Omprof_diff}
\end{figure}

 In Figure~\ref{fig:Omprof_diff}, we plot the distance where the enclosed baryon mass fraction overcomes 90\% of the cosmic baryon mass fraction, $r_{0.9\, f_{\rm b}}$, in units of the virial radius $r_{200}$, as a function of the central value of the halo mass bins considered so far. The colour of every line refers to a different \simba\ variant, and each line style represents a different redshift, as indicated in the legend above the plot. We omitted the No-AGN run, as the results for $r_{0.9\, f_{\rm b}}$ were the same as for the No-jet run.

In general, removing X-ray heating only mildly affects the value of $r_{0.9\, f_{\rm b}}$. On the other hand, if we remove jets at any redshift, $r_{0.9\, f_{\rm b}}$ drops to the virial radius in halos with mass $M_{\rm h} >10^{12.2} \Msun$. By contrast, $r_{0.9\, f_{\rm b}}$ is at least twice as large at $z=2$ and about $\sim 20$ as large at $z=0$ in the fiducial-50 run. Clearly, AGN jets are crucial to evacuate baryons from halos at $z\leq 2$ and $M>10^{12.2}\, \rm M_{\odot}$. This result confirms that found by \cite{Appleby_2021}: \simba\ shows that thanks to AGN jets the missing baryons are not simply undetected, but truly evacuated from halos. 

In the absence of any form of feedback, all halos contain the cosmic share of baryons, at any redshift. In such counter-factual universe, there would be no knee in the bTFR. Stellar and AGN wind feedback (blue lines in  Figure~\ref{fig:Omprof_diff}) have a substantial impact on halos with  $M_{\rm h} < 10^{12.2} \Msun$, as they increase $r_{0.9\, f_{\rm b}}$ by a factor of $\sim 3$ at $z=4$ and up to a factor of $\sim 9$ at $z=0$.

\section{Conclusions and perspectives}
\label{sec:conclusions}

We investigate how baryon-driven physical processes impact the large-scale structure and the internal mass distribution of halos. We did this through the analysis of the \simba\ cosmological simulation, which includes both stellar and AGN-driven feedback prescriptions. We also considered four additional variants of the simulation, where one or more of the feedback mechanisms were turned off. Aside from the role of feedback, we also investigated the impact of the mere presence of baryons on halos and large-scale structure. For this reason, we also ran a dark-matter-only version of \simba.

Throughout our work, we took an outside-in approach. We began by exploring the abundance of different phases of the IGM as a function of redshift in all runs. Then, we characterised the mass distribution across halos by extracting the halo mass function (HMF) and baryon mass function (BMF) at different redshift. Finally, we investigated how different feedback prescriptions alter the mass of individual halos, and the radial density profiles of different baryonic phases (cool/warm/hot CGM gas, ISM, winds, stars) inside halos of different mass, as a function of redshift.

A unified picture for the effect of baryons on all scales emerges from our work: AGN-driven jets are the most important factor for setting baryon contents at $z<2$ and in more massive halos, while stellar feedback dominates at $z>2$ and in lower mass halos. The main findings of our analysis are summarised as follows:

\begin{enumerate}
    \item AGN-driven jets are the main driver for evacuating halos of baryons in massive halos ($\ga 10^{12} \Msun$) at $z<2$ (Figures~\ref{fig:mhalo_diff}, \ref{fig:dprof_diff}). At late times, they are also the main source of heating and cause for the suppression of star formation for such halos (Figure~\ref{fig:mfrac_all}). 
    
    \item In the absence of AGN jets, the baryon mass contained within the virial radius of massive halos is $>90$\% the cosmic baryon mass fraction. If AGN jets are active, one needs to include the baryonic mass up to $10-20$ virial radii away from halos in order to reach the cosmic baryon mass fraction. Hence, the effect of AGN jets is not limited to halos, but reaches out to the IGM (Figure~\ref{fig:Omprof_diff}). Jets drive the amount of hot IGM gas  ($T>10^6 \K$) from $\sim 30\%$ to $70\%$ at $z=0$ (Figures~\ref{fig:fIGM}-\ref{fig:fIGM_feedback}).
    
    \item Stellar feedback is the primary mechanism responsible for the suppression of star formation and gas heating in $\la 10^{12} \Msun$ halos at redshift $z\leq 2$. At higher redshift, the action of AGN jets is minimal, and supernovae-driven winds are the dominant feedback mechanism in all halos (Figure~\ref{fig:mfrac_all}). In contrast, AGN radiative winds have a sub-dominant effect on all statistics considered in this work.
    
    \item The stellar and total baryonic mass of a single halo can vary up to one order of magnitude, depending on which feedback modes are active, and on redshift. In contrast, the total mass of a halo typically varies up to $\sim 30\%$ when all feedback modes are turned off, and less than that if at least one feedback mechanism is active. Therefore, when comparing the effect of feedback on halo properties in different simulations, it is best to select halos by total mass rather than stellar mass (Figure~\ref{fig:mhalo_diff}).
    
    \item The halo mass function is robust to variations of feedback prescriptions, and even to the removal of baryons altogether, within $\sim 20\%$. On the other hand, stellar feedback is crucial in shaping the baryon mass function at $M_{\rm b} < 10^{11} \Msun$, while AGN feedback is the main driver of the suppression of the baryon mass function at the high-mass end (Figures~\ref{fig:hmf_dm}, \ref{fig:mfuncs_diff}).
    
    \item The halo mass function in the fiducial \simba\ $100 \hMpc$ simulation and in the dark-matter-only run do not differ significantly ($\sim 20\%$) at any redshift if the definition of the halo boundaries is changed. However, if the virial radius of a halo is defined such that it contains a mean matter density equal to 2500 times the critical density  of the universe, then the  differences  between the runs with and without baryons can grow up to 50\% at $z=4$ (Figure~\ref{fig:hmf_dm}).
    
    \item The presence of baryons reduces the concentration of dark matter in halos. The DM density is reduced by up to $\sim 20\%$ at $\sim 0.1 \, r_{200}$. Interestingly, the effect of baryons extends outside the virial radius, introducing a $\lesssim 40\%$ increase in the DM density at $\sim 2 \, r_{200}$ in massive ($M_{\rm h} > 10^{12.8}\Msun$) halos (Figure~\ref{fig:dprof_dark}). At high redshift, rapid cooling leads to a concentration of gas and stars towards the halo centre, while at lower redshifts halos become increasingly evacuated of baryons in their outskirts and surroundings (Figure~\ref{fig:Omprof}).

\end{enumerate}

Other exciting avenues for further investigation include a particle-tracing approach to the question of the impact of baryons on halos and large-scale structure. Now that we illuminated what the effect of each feedback prescription is on a wide range of quantities, we can ask ourselves where specific gas particles would end up if certain feedback modules were not active. This would represent an extension of \cite{Borrow_2020} work at different redshift. Furthermore, there is scope for a more in-depth analysis of various aspects of our work, such as the impact of feedback on the concentration and shape of halos, on the topology of the cosmic web and on the spatial correlations of the matter density field. We hope to address these questions in forthcoming work.

\section*{Acknowledgements}

We are grateful to John Peacock and the members of the \simba\ collaboration for helpful discussions. We acknowledge the \texttt{yt} team for development and support of \texttt{yt}. Throughout this work, DS was supported by the European Research Council, under grant no. 670193, by the STFC consolidated grant no. RA5496, and by the Swiss National Science Foundation (SNSF) Professorship grant no. 202671. RD acknowledges support from the Wolfson Research Merit Award program of the U.K. Royal Society. WC is supported by the STFC AGP Grant ST/V000594/1 and the Atracci\'{o}n de Talento Contract no. 2020-T1/TIC-19882 granted by the Comunidad de Madrid in Spain. He further acknowledges the science research grants from the China Manned Space Project with NO. CMS-CSST-2021-A01 and CMS-CSST-2021-B01. SA is supported by a Science \& Technology Facilities Council (STFC) studentship through the Scottish Data-Intensive Science Triangle (ScotDIST). This work used the DiRAC\MVAt Durham facility managed by the Institute for Computational Cosmology on behalf of the STFC DiRAC HPC Facility. The equipment was funded by BEIS capital funding via STFC capital grants ST/P002293/1, ST/R002371/1 and ST/S002502/1, Durham University and STFC operations grant ST/R000832/1. DiRAC is part of the National e-Infrastructure. 
This work made extensive use of the NASA Astrophysics Data System and of the astro-ph preprint archive at arXiv.org. %For the purpose of open access, the author has applied a Creative Commons Attribution (CC BY) licence to any Author Accepted Manuscript version arising from this submission.\\

\section*{Data availability}

The simulation data underlying this article are publicly available\footnote{\url{http://simba.roe.ac.uk}}. The software used in this work is available on Github\footnote{\url{https://github.com/danieleTS/baryons_simba}} and the derived data will be shared upon reasonable request to the corresponding author.

%%%%%%%%%%%%%%%%%%%%%%%%%%%%%%%%%%%%%%%%%%%%%%%%%%



\bibliographystyle{mnras}
\bibliography{lss_baryons} 

%%%%%%%%%%%%%%%%%%%%%%%%%%%%%%%%%%%%%%%%%%%%%%%%%%

%%%%%%%%%%%%%%%%% APPENDICES %%%%%%%%%%%%%%%%%%%%%

\appendix

\section{Convergence tests}
\label{sec:convergence}

In this appendix, we present the convergence tests relevant to the results examined in this work. To do this, we consider the simulations where all the feedback prescriptions are active, with box sizes $25 \hMpc$, $50 \hMpc$ and $100 \hMpc$. This allows us to check whether our results are converged volume wise. To verify the convergence with respect to mass resolution, we also consider the \simba\ High-res run (see Table~\ref{tab:runs}).

\subsection{Mass functions}

\begin{figure}
	\includegraphics[width=\columnwidth]{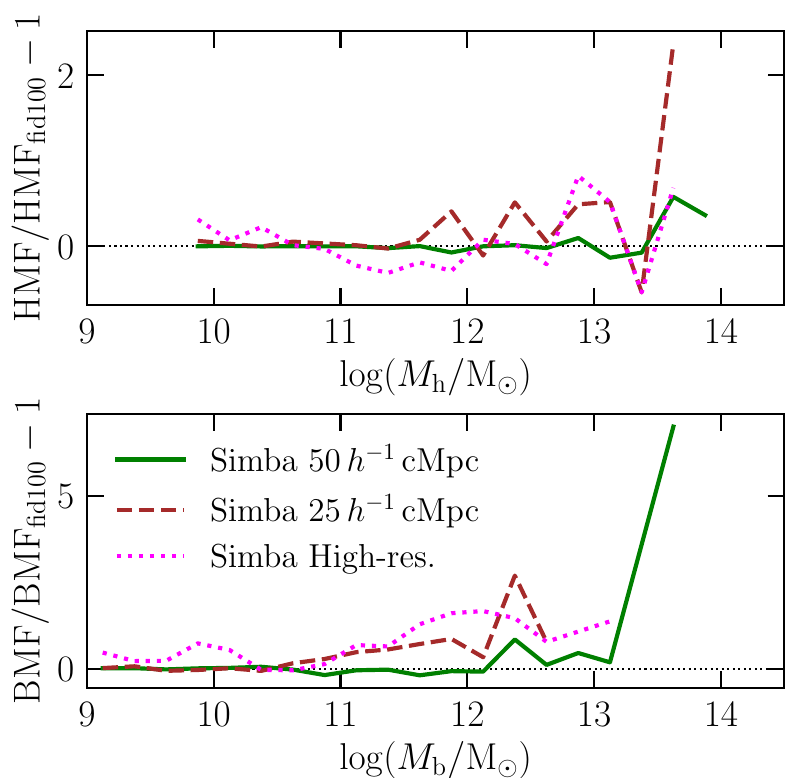}
    \caption{Convergence test for the halo and baryon mass functions (upper and lower panel, respectively). The solid green, dashed brown and dotted magenta lines refer to the relative difference of the mass functions in the \simba\ $50 \hMpc$, \simba\ $25 \hMpc$ and \simba\ High-res results, respectively, relative to the \simba\ $100 \hMpc$ simulation. All results are obtained at redshift $z=0$.}
    \label{fig:mf_conv}
\end{figure}

We begin by testing the convergence for the mass functions. The upper panel of Figure~\ref{fig:mf_conv} shows the relative difference between the HMF of the aforementioned runs with respect to the \simba\ $100 \hMpc$ simulation at $z=0$. The solid green, dashed brown and dotted magenta lines refer to the \simba\ $50 \hMpc$, \simba $25 \hMpc$, and \simba\ High-res runs, respectively. The horizontal dotted line marks the zero difference level, and serves as a guide. 

We notice that runs with same mass resolution but different volumes exhibit relative differences within $\sim 10\%$ up to $M_{\rm h} \approx 10^{12} \Msun$, and within $\sim 50\%$ up to $M_{\rm h} \approx 10^{13} \Msun$. However, at larger halo masses the mass functions can differ up to a factor of $\sim 3$. If we consider the \simba\ High-res run, we can see that its predictions for the HMF match those of the \simba\ $100 \hMpc$ simulation within $\sim 20\%$ up to $M_{\rm h} \approx 10^{12} \Msun$. For more massive halos, the mass functions can exhibit differences up to a factor of $\sim 2$ (at $M\approx 10^{13} \Msun$).

The lower panel of Figure~\ref{fig:mf_conv} shows the relative variations in the BMF across difference runs, with the same colour coding as in the upper panel. The BMFs exhibit comparatively larger differences: for halos with $M_{\rm b} \lesssim 10^{12.5} \Msun$ these are within a factor of $\sim 2$ when we change the box size, and up to a factor of $\sim 3$ if we increase the mass resolution. For larger baryon masses, the BMF given by the \simba\ High-res run can exceed the predictions of the \simba\ $100 \hMpc$ by a factor of $\sim 8$. Also, reducing the volume from $100 \hMpc$ to $50 \hMpc$ causes the BMF to increase by a factor of $\sim 6$ for $M_{\rm b} \gtrsim 10^{13.5} \Msun$.

In conclusion, both the HMF and BMF are well converged both resolution-wise and volume-wise except at the high-mass end. The larger discrepancies at the high mass end are expected because of the lower number of massive halos. Overall, convergence is tighter for the HMF than for the BMF. This is likely due to the fact that the HMF is based on the total mass of the halo, and hence all particle types within the halos count for the HMF. On the other hand, the BMF depends only on the baryon particles within halos. These make up only a fraction of the total halo mass, and it is easier to see effects of the finite mass resolution of the simulations if the number of particles is smaller. We verified that our overall conclusions on the convergence of the HMF and BMF are unchanged for the other snapshots considered in this work ($z=1$, $z=2$, $z=4$).

\subsection{Density profiles}

\begin{figure*}
    \centering
    \includegraphics[width=\textwidth]{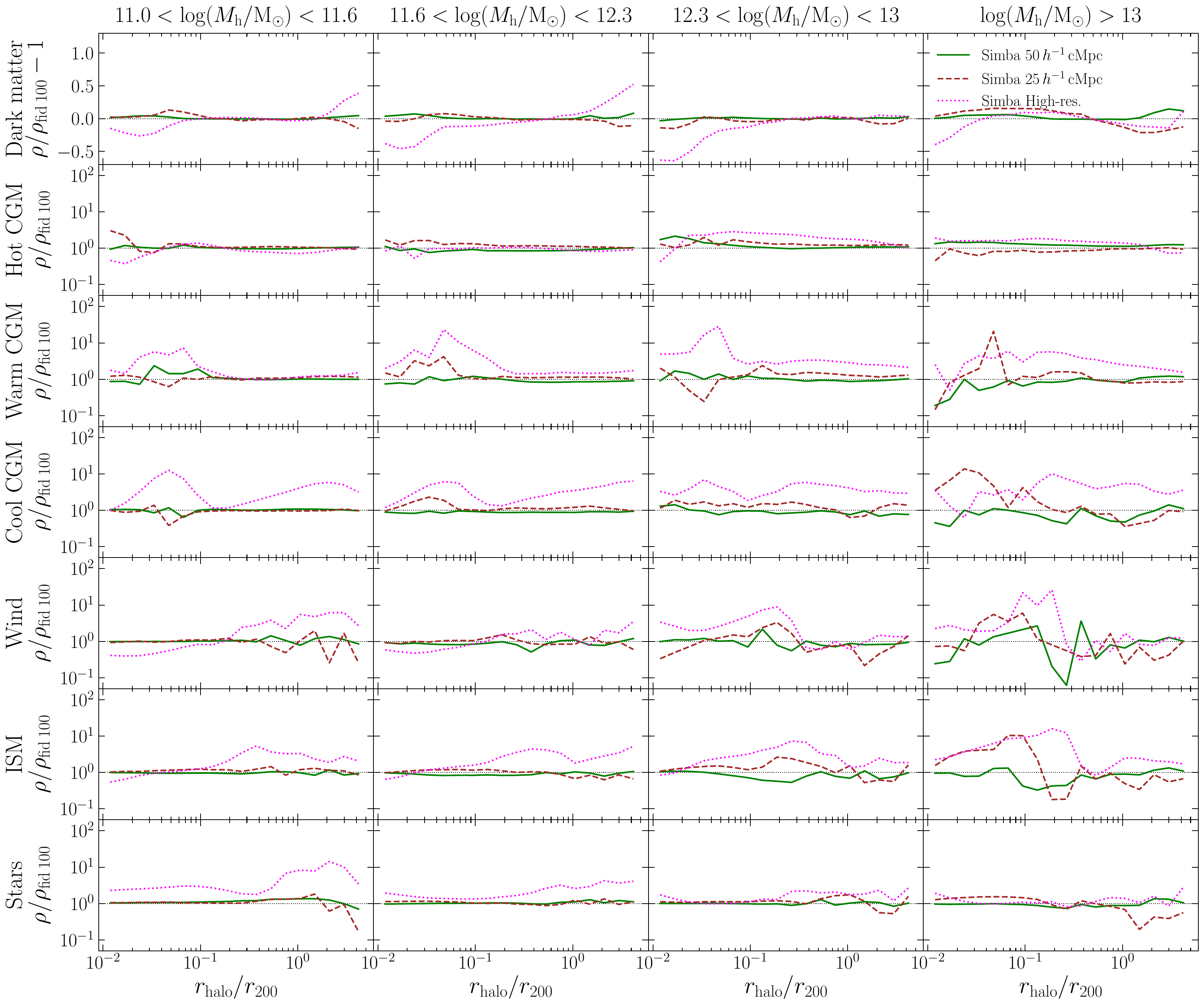}
    \caption{Convergence test for the density profiles within halos. The panels are organised in the same way as in Figure~\ref{fig:dprof_diff}. The solid green, dashed brown and dotted magenta lines refer to the relative differences (for the DM profile only) or ratio (for all other components) of the profiles in the \simba\ $50 \hMpc$, \simba\ $25 \hMpc$ and \simba\ High-res, respectively, with respect to the \simba\ $100 \hMpc$ simulation. All results are obtained at redshift $z=0$.}
    \label{fig:dprof_conv}
\end{figure*}

We now  test the convergence of the density profiles at. We will show our results for $z=0$, but we verified that we obtain qualitatively similar results at higher redshift too. Figure~\ref{fig:dprof_conv} has the same structure as Figure~\ref{fig:dprof_diff}: every row corresponds to a different component that makes up halos, and every column represents a different total halo mass bin. In the first row, we plot the relative difference of the DM mean density profile obtained with different runs, with respect to the fiducial-100 run, following the same colour-coding as in Figure~\ref{fig:mf_conv}. In the other rows, we show the ratio between the mean density profiles of the component, specified in the left part of the figure, given by the various runs, with respect to the \simba\ $100 \hMpc$ simulation.

Overall, for $r>0.1\, r_{200}$ the convergence in the density profiles is good both with respect to mass resolution and volume, expected from halo-to-halo scatter. For $r<0.1\, r_{200}$, the density profiles are still generally converged volume-wise, but often not with respect to the mass resolution. This holds also for $r<0.1\, r_{200}$ in most cases. However, for some components (warm and cool CGM, ISM, wind and stars) and mostly in high-mass haloes ($M_{\rm h} >10^{12.2} \Msun$, volume-wise convergence is weaker, as runs with different volumes but same resolution can differ up to an order of magnitude. The convergence in mass resolution tends to be better in the highest-mass halos, where smaller radial bins contain more particles than in lower-mass halos. However, the stellar density profile is not optimally resolved resolution-wise in the lowest mass bin. Full convergence is not achieved for the ISM density profile either, indicating that the criterion to define ISM gas may be particularly sensitive to resolution.

In conclusion, we believe that our results for the density profiles are generally reliable for both outside and within $r< 0.1\, r_{200}$. Although convergence with respect to mass is still achieved in the latter regime for some of the halo components considered, we believe that it would be necessary to run higher-resolution simulations with large volume in order to obtain truly robust results. This is particularly the case for the stellar and ISM density profiles, for which the convergence with respect to resolution is not optimal. We leave this for future work.

%%%%%%%%%%%%%%%%%%%%%%%%%%%%%%%%%%%%%%%%%%%%%%%%%%

% Don't change these lines
\bsp	% typesetting comment
\label{lastpage}
\end{document}